\documentclass[aps,prx,twocolumn,groupedaddress]{revtex4-2}
\usepackage{graphicx}
\usepackage{xcolor}
\usepackage{amsmath}
\usepackage{tikz}
\usepackage[percent]{overpic}
\usepackage{hyperref}
 
\begin{document}

\title{Activity enhances heterogenous dynamics in 2D granular glasses}

\author{Chetan Yadav and Narayanan Menon}

\affiliation{Department of Physics, University of Massachusetts
Amherst, Massachusetts 01003, USA}

\date{\today}

\begin{abstract}
 We study the effect of self-propulsion in the approach to the glassy regime of 2D  bidisperse mixtures of frictional discs, confined to a horizontal plane, and fluidized by vertical vibrations. The discs are designed either to be `passive’, that is, exhibiting isotropic motion in the plane, or `active’, with a preferred polar mobility. When dynamics in mixtures of large active and small passive discs are compared to mixtures of large and small passive discs, the active system shows faster dynamics at equal area fractions. To explore whether activity merely speeds up motion, we select active and passive mixtures with the same relaxation time and compare their dynamics.  Many indicators show that dynamical heterogeneity is much greater in the active glass with longer-ranged spatial velocity correlations, larger temporal variations in displacement, and higher amplitude of the four-point correlator. In this very crowded regime, despite frequent collisions, the activity directions of the discs remain persistent. However, the activity direction of the discs remains spatially uncorrelated, and furthermore, the particle displacements are not aligned with the activity direction, thus presenting a puzzle as to how activity leads to such strong heterogeneity.

\end{abstract}

\maketitle

\section{Introduction}

The puzzle of the glass transition is that the dynamics of a liquid becomes extremely sluggish, without a concomitant change in structure. In this state, any structural relaxation requires complicated cooperative rearrangements involving several particles.  Conventional glasses are born from equilibrium dynamics, with particles connected to a thermal bath and cooled or compressed fast enough to suppress crystallization.  Recent simulations consider adding a nonequilibrium ingredient to this problem by mixing in with these `passive' particles some fraction of `active' particles \cite{Berthier2019JCP, Janssen2019IOP}. Both the passive and active particles are connected to a heat bath, but the active particles additionally experience a persistent directed force or a velocity whose orientation diffuses over a longer timescale $\tau_\theta$ . 

We might guess that the introduction of these active particles into the glass helps fluidize the system and speed up relaxation. This intuition is confirmed by a number of simulations on active glasses \cite{Mandal2016SM,Paul2023PNAS,Keta2022PRL,Berthier2017NJP,Sollich2023extreme}.  These simulations collectively explore a range of parameters and 
several variants of active Ornstein-Uhlenbeck  (AOUP) models \cite{Martin2021PRE}, in both 2D and 3D systems. Some simulations explore the limit of long persistence times, whereas others operate over a range of $\tau_\theta$. 

 Deep in the glassy state, traditional glass formers (or passive glasses) slow down heterogeneously, with growing spatial variations in relaxation dynamics that persist on the relaxation time scale \cite{ediger2000spatially, Kob1997PRL}. It is natural to ask whether the introduction of activity has an effect on these dynamical heterogeneities. 

In addition to the speeding up of the relaxation, it is robustly observed in all the simulated systems that  dynamical heterogeneity is greatly enhanced by the presence of active particles, relative to a purely passive glass. Particle displacements are more burst-like, velocity distributions become strongly non-Maxwellian, and the range and amplitude of velocity correlations grow as the persistence time and area fraction are increased \cite{Keta2022PRL}.  Higher order correlations functions ($\chi_4$) show stronger peaks in active glasses than in their passive counterparts \cite{Paul2023PNAS}.

Experimental exploration of the same questions have been lacking despite important differences in the microscopic ingredients in experimental systems and these simulations. The simulations consider frictionless systems where the orientation dynamics are unaffected by collisions and only relax due to coupling with a noise term. That is, the orientational relaxation time is a control parameter that is held fixed in the simulation. However, in our active granular systems, interparticle interactions are frictional. As a result, collisions affect orientational degrees of freedom, and the persistence time of the activity direction changes with volume fraction through the collision rate. Thus, the basic parameter that specifies the level of activity in a simulation is an emergent quantity in the experiment, rather than a control variable.  A further difference is that, in typical experimental situations, the single particle persistence length can be quite small, of the order of a few particle diameters, whereas simulations often explore extremely high persistence situations, because the effects of activity are presumably most pronounced in this limit.     

Other mechanisms that couple the rotational and translation dynamics include self-alignment \cite{Baconnier2025RMP},  the property of a polar active unit to align or anti-align its activity orientation and its velocity. A coupling between rotation and number density interaction has also been  constructed by Zhang \textit{et al}.\cite{Zhang2021NP} where they fabricate asymmetric particles subject to torques that reorient particle motion towards high-density regions.  Anisotropic particle shape also leads to gain reorientations under collisions, even without friction.  However, shape effects lead to stronger geometrically-driven tendency to align and dynamics influenced by ordering defects (see e.g. \cite{narayan2007long},\cite{Arora2022PRL}). 

A distinct dissipative mechanism in dry granular systems is inelasticity, which leads to damping of the relative particle motion at high densities. This effect has long been recognized to induce velocity correlations [e.g.\cite{blairkudrolli2001,soto2001statistical}] and can have consequences for structure \cite{van1999randomly}, phases \cite{Komatsu2015PRX}, and collective dynamics \cite{Chen2024NatComm}. 

These inelastic effects make it vital for us to make parallel experiments with passive and active granular systems in order to cleave apart the effects of inelasticity and activity.

Beyond granular systems, the dense active glass has also been explored in the dynamics of dense human crowds \cite{Gu2025Nature}, and in the glass-like dynamics of collective cell migration \cite{Angelini2011PNAS}, which demonstrate analogies between cell monolayer 
migration and particulate glass formers. 

 In this paper, we report on experiments with two bidisperse mixtures of granular discs serving as model 2D glass formers. One of these is comprised of small and large passive particles, and the other of small passive and large active particles. We show the effects of activity, by comparing these two systems as their area fractions, $\phi$, are increased. The mixture with active particles shows faster dynamics when compared at equal area fraction. To isolate effects that go beyond this speedup in the dynamics, we compare pairs of active and passive mixtures that are chosen to have the same relaxation time by selecting slightly different area fractions.  We find by several measures of the dynamics that the active system has much more pronounced dynamical heterogeneity in both time and space. 
 Since both systems are frictional, their rotational dynamics are coupled to interparticle collisions.  As a result, the rotational relaxation time $\tau_\theta$ of the discs increases with increasing area fraction, so the active direction of the active discs remains persistent in the glassy regime.  However, in this crowded environment, the alignment between the activity and displacement vectors is lost. Thus, the strong qualitative effect of activity in the glassy regime comes despite the absence of any spatial organization of the  activity vectors and any obvious motion in the local activity direction.
 
\section{Experimental Setup}

\begin{figure*}
    \centering
    \includegraphics[width=1.0\linewidth]{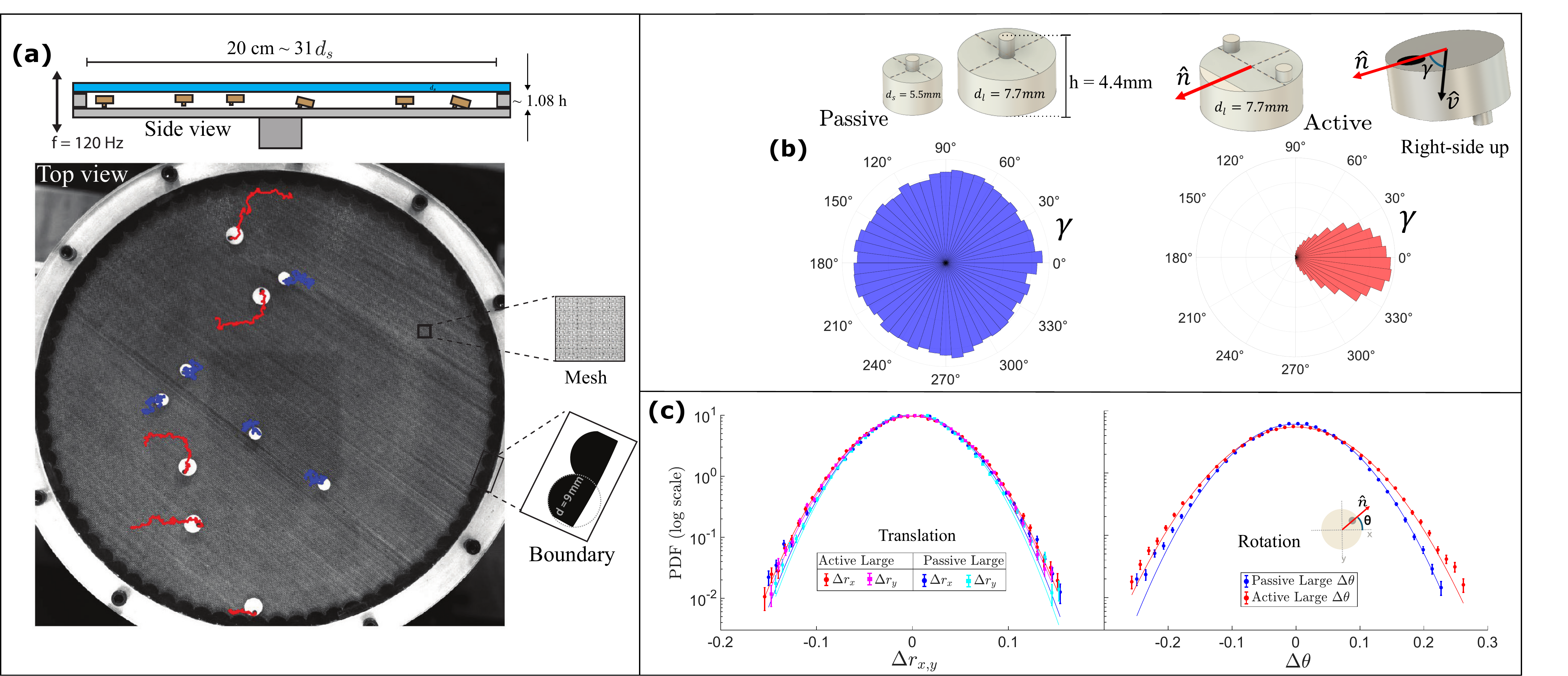}
    \caption{Experimental setup (a) Side view of the cell:  The gap between the bottom plate and top plate is 1.08 times the height of the large particle. Top view of the cell:  The diameter of the cell is 31 $d_S$ . Active particles are shown with red tracks, and passive particles with blue tracks for 192 shakes (200 frames). The bottom plate is covered with mesh to suppresses small asymmetries in leveling or vibration. 
    The bumpy boundary suppresses global system rotation and crystalline packing at the walls at high density. (b) Particle design: Both small and large passive particles (diameter $d_s$ and $d_l$) are discs with a pin at the center. The active particles are discs with two off-center pins. All particles are decorated with an off-center dot to determine orientation $\theta$ relative to a lab-fixed frame. In the case of the active particle, this also defines the mobility direction via $\hat{n}.\hat{x} = \cos\theta$.
    Polar histogram of angle $\gamma$ between displacement and $\hat v$ (computed over time $\tau_{\theta}$) and activity direction $\hat{n}$: Left: passive particle has isotropic distribution; Right: active particle is polar, with a preferred direction of displacement. 
    (c)  Distributions of translational  (left) and orientational (right) displacement of an isolated particle following a single shake. The distribution in each case is nearly gaussian and are quantitatively similar for active and passive discs.
    \label{fig:exp}}
\end{figure*}

\subsection{Particles, cell geometry and shaking}
We study a vibrated bidisperse mixture of equal numbers of small and large disks with diameters $d_s$ and $d_l$, respectively, in the size ratio of 1:1.4, which is well known from hard disk simulations \cite{Speedy1999JCP} to be a stable glass-forming formulation.  

\textbf{Cell geometry}  As shown in Fig. \ref{fig:exp}(a), the system of particles is placed in a circular geometry of diameter $31 d_s$, and is confined to quasi-2D motion with a transparent lid.  The gap between particle and the lid is  $\approx 8\%$ of the particle height so that discs remain in a monolayer. A nylon mesh with opening size much smaller than $d_s$ is glued to the floor to suppress the tendency to slip in response to any slight tilt.
We use a scalloped boundary with small bumps (on the scale of the particle size) to suppress ordering at the boundary and any tendency for bulk rotation at high packing fractions.  The bumps are also relatively small, so that they do not generate flows.

 \textbf{Particles} 
As shown in Fig. \ref{fig:exp}(b) the particles are 3D printed with diameters of $d_s = 5.5$ and $d_l = 7.7$ mm for the small and large discs, respectively. The smaller discs are decorated with a cylindrical pin at the centre, endowing them with isotropic motion in the horizontal plane under vibration. 
We refer to these as `passive' particles. The larger discs are designed to be either similarly passive, with one central pin, or they are rendered `active' by placing two cylindrical pins displaced from a diameter.  As shown, the pins define a body axis for the active particles along which they show polar motion.

\textbf{Vibration} The system is fluidized by vertical vibration using an electromagnetic shaker (LDS456) coupled to the cell by a slender rod that passes through a square air bearing that constrains rotation and horizontal motion. For the studies reported here, the frequency is held fixed at $f=120$ Hz, and the signal amplitude is set to produce an acceleration of $\Gamma = A\omega^2 = 7.5 g $. The amplitude of the shaking is $\approx 3\%$ of the particle height. 

\textbf{Imaging}  We image the kinematics at $125$ frames/s, slightly different from the shaking frequency $f$. The position of the particles is determined by finding the centroids of all discs. The orientation $\theta$ relative to a lab-fixed frame is located by an off-centre dot. The centroid-to-dot vector, $\hat{n}$, defines the body orientation for all particles, which is also the polarity vector for the active particle.  Time series of position and orientation for all the particles are acquired for a duration of  15360 $f^{-1}$, or shakes.

\textbf{Area fraction}
 We have calculated the occupied area fraction $\phi$ from the particle count multiplied by the detected particle area, and the total area accessible to the particles (more details are in Appendix \ref{app:area_fraction}). Due to the finite system size there is wall-induced structure so the outermost layer of particles is excluded from the analysis. In the limit of infinite system size $\phi_{rcp} = 0.846$ for our bidisperse mixture. However, there is an intrinsic lowering of the jammed area fraction due to finite size, which has been carefully characterized in previous simulation studies \cite{Desmond2009PRE}. The maximum area fraction we are practically able to study is $\phi = 0.823$.

\textbf{Single particle motion}  To compare the short time dynamics of the active and passive discs we show in Fig.\ref{fig:exp}(c) the probability distributions of their rotational and translational displacements at a time scale $\Delta t = f^{-1} = 1$ shake.  The short-time displacements are close to gaussian and the distributions for the large active and passive  discs are very similar to each other, both for translational and rotational kicks. In other words, the kinetic granular temperatures are comparable for the active and passive large discs (Appendix Table~\ref{tab:displacements}) \cite{Chen2024NatComm}. 

Despite the similarity in short-term dynamics, the long-term trajectories of active and passive particles (blue and red traces, respectively) are very different as seen in Fig.\ref{fig:exp}(a). The displacement statistics of the passive discs are brownian, whereas the active discs show persistent bias for motion in the $\hat{n}$ direction. 
The dynamics of the active discs are fit to the active brownian particle model \cite{Walsh2017SM}. We extract a persistence length $l_p$ over which a particle moves before its direction is randomized by calculating the orientation-displacement correlation (Appendix - Fig.\ref{fig:OD_corr}), and determine $l_p \approx 4 d_s$.  
The timescale $\tau_\theta$ over which an initial orientation $\hat{n}$ wanders is specified by the time at which the mean square orientation becomes unity, namely,  $\langle(\theta(t+\tau_\theta)-\theta(t))^2\rangle = 1$. Here, the average $\langle\rangle$ is taken over both particles and initial time $t$. For isolated active particles, we find $\tau_\theta^0= 94  \text{ shakes}$. (This is similar in spirit to the decay time $D_R^{-1}$ of the orientation-orientation correlation function, but is subtly different quantitatively) (Appendix \ref{app:rotation}). 

Another useful measure of polarity is found in the activity-velocity alignment by calculating the angle $\gamma$ between the body axis direction $\hat\eta(t)$ and velocity $\hat v(t+\Delta t)$, where we choose $\Delta t=\tau_\theta$. Once again, the contrast between the isotropic dynamics of the passive particle and the biased motion of the active particle is shown in Fig.\ref{fig:exp}(b) in the form of polar histogram.
 
\section{Results}

We present dynamics and structures of binary mixtures of (1) large  and small passive particles, and (2) large active and small passive particles, which we refer to as passive and active mixtures, respectively.  The driving parameters ($f$ and $\Gamma$) are held fixed, and the glassy regime is approached by increasing the total area fraction $\phi$ covered by the particles. At each area fraction we vibrate the system long enough to equilibrate the configuration, and then increase $\phi$ by adding particles. Thus the data are taken in one sequence to avoid any possible residual dependence on initial configurations. 
\subsection{Approach to the glass transition}

\begin{figure}
    \centering
    \includegraphics[width=1.0\linewidth]{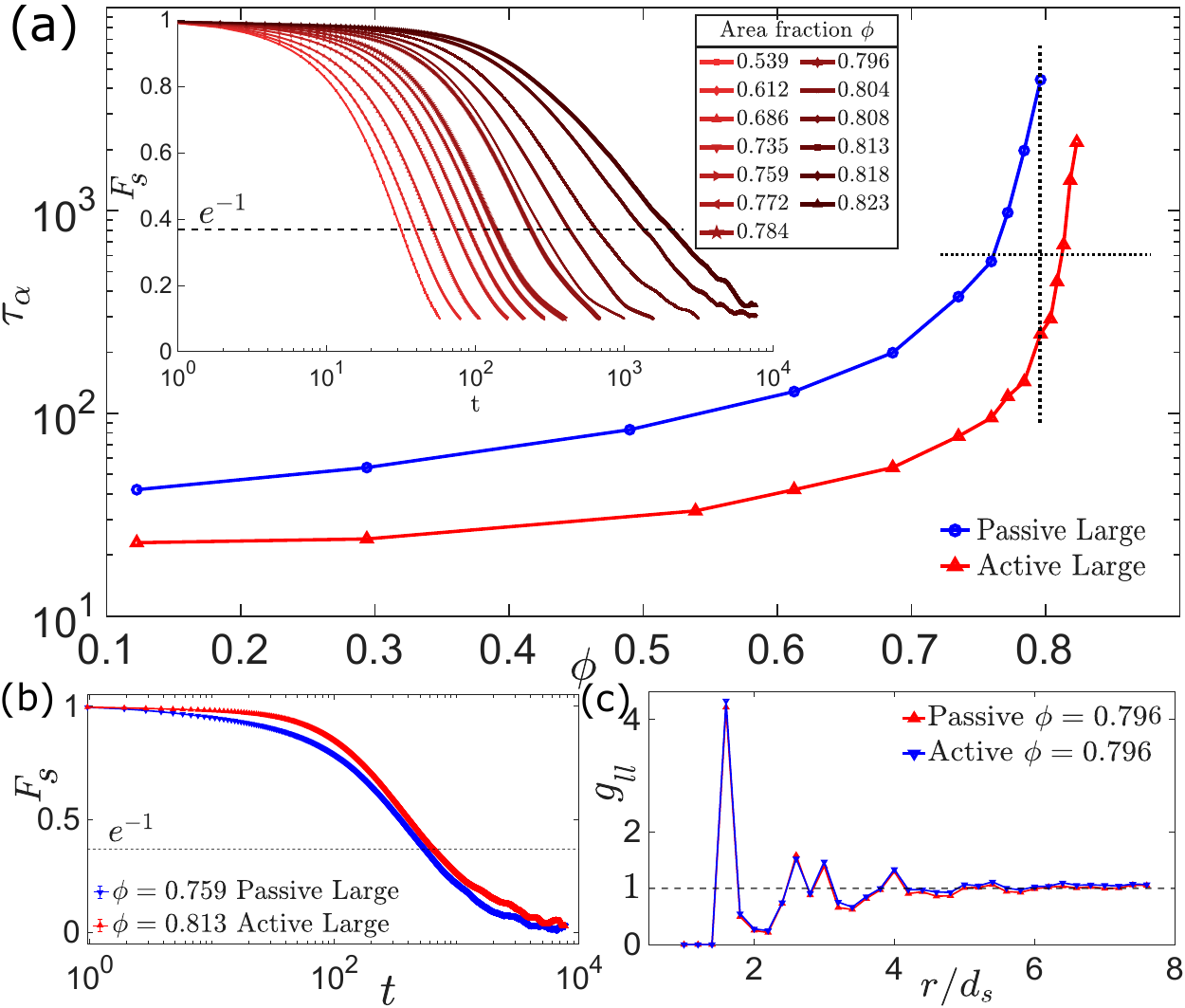}
        \caption{\textbf{(a)} Relaxation time $\tau_\alpha$ vs. area fraction $\phi$ for the large discs in the active and passive mixtures, showing a sharp increase as the glass is approached. 
    The relaxation time of the active mixture (red) is shorter than the passive mixture (blue) when compared at the same $\phi$.  The relaxation time is obtained from the self-intermediate $F_S(q,t)$, which is shown in the inset for the active mixture for several values of $\phi$. $\tau_\alpha$ is chosen to be the time at which $F_s$ decays to  $e^{-1}$. \textbf{(b)} Examples of a pair where $F_s$ for the active system and passive mixtures have the same translation relaxation time at different area fraction. This pair is indicated in (a)  with a horizontal dashed line. \textbf{(c)} Radial distribution function for a pair of active and passive mixtures at the same $\phi$. This structural measure is identical for these two mixtures which have very different $\tau_\alpha$, as shown by the vertical dashed line in (a).
    \label{fig:rlx}}
\end{figure}

Our first characterization of the slowing translational dynamics in the two mixtures is by the conventional method of calculating the self-intermediate scattering function. 
\begin{equation}
F_S(q;t) = \frac{1}{N} \left\langle \sum_{n} e^{i \mathbf{q} \cdot [\mathbf{r}_n(t) - \mathbf{r}_n(0)]} \right\rangle,
\end{equation}
where $q = 1/d_s$, $d_s$ is the small particle diameter and $n$ runs over all the large particles. The angled bracket represents an average over initial time $(t=0)$. 
As shown in the inset to Fig. \ref{fig:rlx}a, we define relaxation time $\tau_\alpha$ as the time it takes for $F_S(q;t)$ to decay to $1/e$, that is, $F_S(q,\tau_\alpha) = e^{-1}$. 
This is a measure of the timescale for particles to move by a distance of the order of $d_s$. 

 In both the active and passive mixtures, the relaxation time $\tau_{\alpha}$ increases with increasing area fraction $\phi$ as shown in the main Fig.\ref{fig:rlx}(a)). The relaxation time increases  faster than  exponentially, especially as we approach the glassy regime.
 At a given area fraction, the structure of the mixture as measured by the radial distribution function $g(r)$ is the same, as shown in Fig.\ref{fig:rlx}(c).  However, at all area fractions, the relaxation in the active mixture is faster than the passive mixture, with the red curve always sitting below the blue curve in Fig.\ref{fig:rlx}(a). Thus, active particles  speed up relaxation in a mixture with no change in the static structure. 

 In many of the comparisons to follow in this article, we thus compare the dynamics of the active and passive mixture at the same relaxation time, $\tau_{\alpha}$ (Fig.\ref{fig:rlx}(b)) rather than at the same area fraction $\phi$, in order to identify qualitative differences in these systems that go beyond the speeding up of the dynamics by activity.

\begin{figure}
    \centering
    \includegraphics[width=1.0\linewidth]{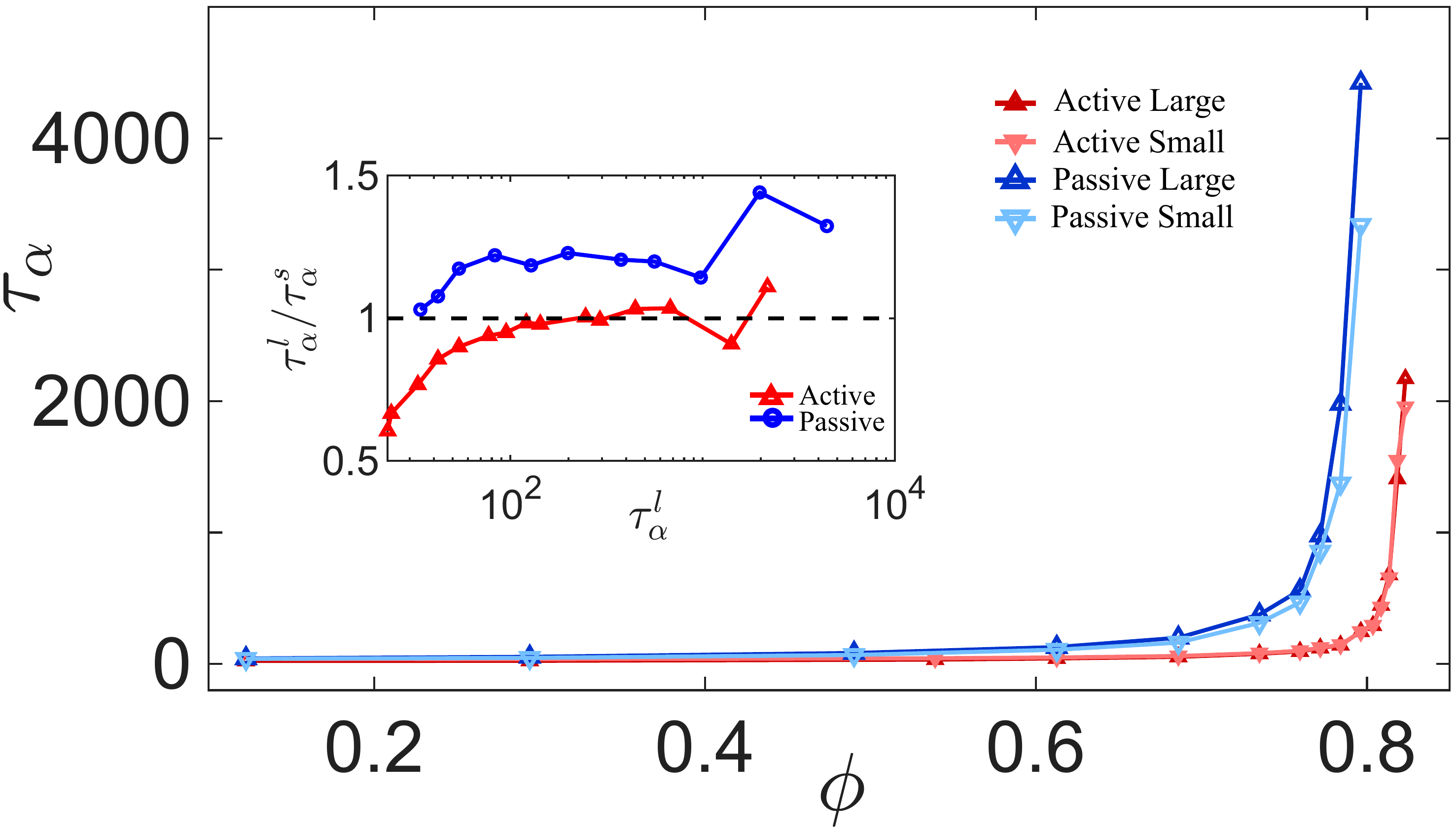}
    \caption{Comparing relaxation time $\tau_\alpha$ of large and small particles as function of area fraction $\phi$. In the active system the small and large particles share the same relaxation time, whereas in the passive system small particles have faster dynamics. Inset: $\tau_\alpha^l/\tau_\alpha^s$ approaches 1 in the active system. }
    \label{fig:rlx_both}
\end{figure}

In Fig.\ref{fig:rlx_both}, we show a qualitative difference between the mixtures even at the level of the mean relaxation time. We extract separately the relaxation times $\tau_\alpha^l$ and $\tau_\alpha^s$ for the large and small particles, respectively, and find that in the passive mixture, small particles rearrange faster than large particles at large $\phi$. On the other hand, in the active mixture, the small and large particles are  equilibrated in the glassy regime, suggesting larger scale cooperative motion in which both types of particles get entrained. 

\begin{figure}
    \includegraphics[width=1.0\linewidth]{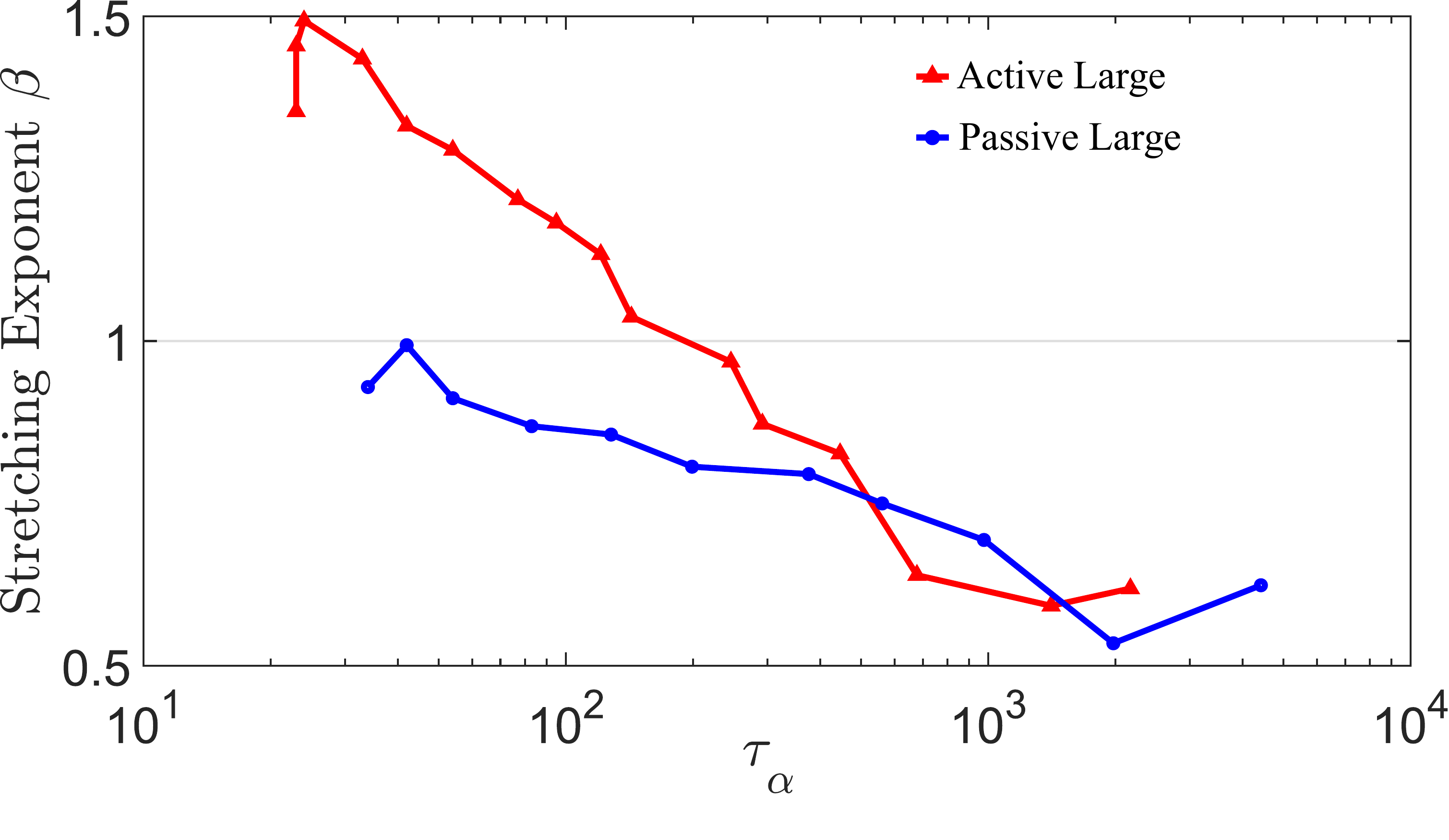}
    \caption{The width of the relaxation is characterized by fitting the self-intermediate scattering function $F_s(q,t)$ to a stretched exponential $e^{-(t/\tau)^\beta}$. $\beta$ decreases  as $\tau_\alpha$ increases, signaling a broader relaxation spectrum in the glassy regime. In the active mixture, dynamics away from the glass (low $\phi$) are fit by a `compressed' exponential ($\beta > 1$).}
    \label{fig:beta}
\end{figure}

 In conventional glasses, the rapid slowing down of dynamics is also accompanied by a broadening of the shape of the relaxation. We parametrize the shape of the self-intermediate scattering function by fitting to the well-known stretched exponential function: $F_S(q,t)=exp(-(t/\tau_{\alpha})^\beta)$. For a conventional glass former, the relaxation is exponential ($\beta=1$) at low densities or high temperatures, with $\beta$ decreasing as the glass transition is approached and values near 0.5 are often observed near glass transition for fragile supercooled liquids \cite{Ediger1996ACS}. This is exactly the behaviour we observe in our passive system, as shown in Fig.\ref{fig:beta} (blue symbols).  However, the active system at low densities shows $\beta>1$ , as might be expected for dynamics that are faster than exponential, crossing over to a stretched exponential only in the very dense regime. This is true for both the species i.e. large and small (Supplementary Material \cite{Supplementary} Fig.\ref{fig:beta_both}).
 The broadened relaxation has been interpreted as an inhomogeneous average over regions of  relaxation dynamics with varying relaxation times, however, the observations of a `compressed' exponential crossing over into a stretched exponential challenge that interpretation and imply correlated motion rather than a superposition of exponential dynamics.   

\subsection{Active mixture is more dynamically heterogeneous}

Dynamical heterogeneity has become one of the defining features of the glassy regime \cite{Berthier2011Physics}.  
In the previous section, we averaged over the fluctuations in the two-point correlations. A more insightful quantification of these fluctuations is obtained by calculating a four-point susceptibility \cite{CDasgupta1991EL,Kob1997PRL}. The general result is dominated by terms that pick out the self contributions \cite{Keys2007NP,Glotzer2000JCP}: 

\begin{equation}
\begin{split}
\chi_{4}(t,a) &= N \left[ \langle Q^{2}(t,a) \rangle - \langle Q(t,a) \rangle^{2} \right], \\
Q(t,a) &= \frac{1}{N} \sum_{i} \Theta \!\left( a - \left| \mathbf{r}_{i}(t) - \mathbf{r}_{i}(0) \right| \right). \\
\end{split}
\end{equation}
where $a = 0.1d_s$.

\begin{figure}[tb]
    \centering
    \includegraphics[width=1.0\linewidth]{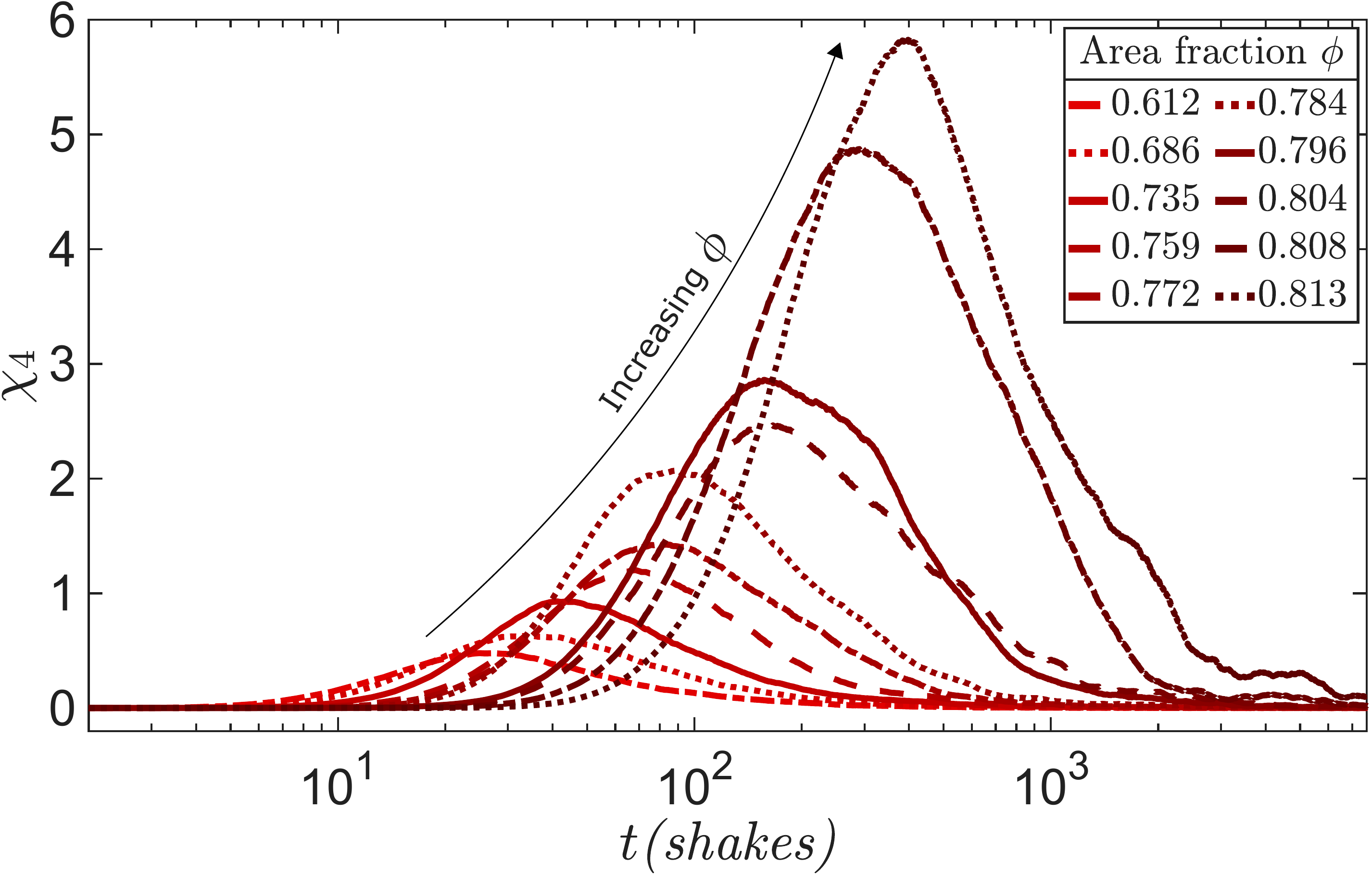}
    \caption{ The time-dependent four-point susceptibility, $\chi_4(t)$ shown here  for  large particles in the active mixture. The dynamical heterogeneity increases strongly towards the glassy regime, as signalled by the increase in amplitude and peak position of $\chi_4(t)$ with increasing $\phi$.}
    \label{fig:x4_al}
\end{figure}

Thus defined, $\chi_4$ is the  variance of the fraction $Q(t,a)$ of particles which have moved a threshold distance $a$ over a time $t$. Even in a heterogeneous system the variance is zero at short  and long times since very few, and almost all, of the particles have moved by $a$, respectively. At  intermediate time there is a much bigger spatial-temporal variation in this quantity, reflecting heterogeneity in the dynamics. Figure \ref{fig:x4_al} demonstrates that as the area fraction $\phi$ is increased, there is a monotonic increase in height of the peak reflecting an increasing amplitude of heterogeneity, the peak of  $\chi_4$ shifts to the longer time scales, and the curve starts to widen showing that the system is heterogeneous at a broader range of timescales.

\begin{figure}[h!]
  \centering
  \includegraphics[width=1.0\linewidth]{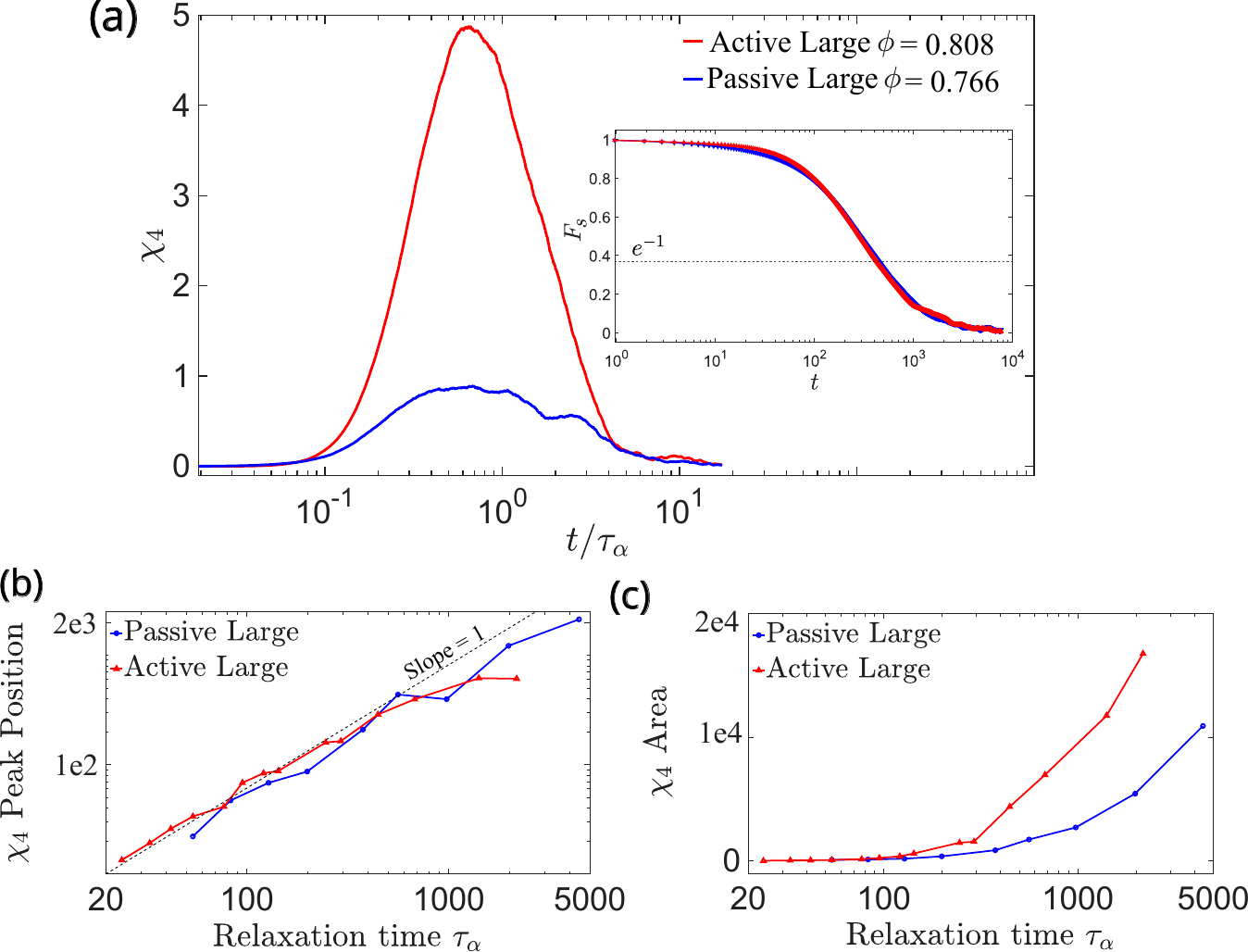}
  \caption{(a) Comparing dynamical heterogeneity via $\chi_4(t)$  for a pair of active and passive mixtures with the same relaxation time, as shown in the inset. They peak at the same timescale, slightly below $\tau_\alpha$, but the active system is more heterogeneous.
    (b) The peak in $\chi_4(t)$  in the two mixtures tracks  the relaxation time $\tau_\alpha$, as area fraction $\phi$ is changed.
    (c) Magnitude of heterogeneity, measured as the area under the $\chi_4$ curve, is always higher in the active system, with the gap increasing at higher $\phi$.}
  \label{fig:x4_combined}
\end{figure}

In Figure \ref{fig:x4_combined} we compare these trends for the active and passive mixtures. An example of an active and passive system with the same $\tau_\alpha$ are shown in Fig.\ref{fig:x4_combined}a: despite very similar two-point correlations (inset), the peak amplitude of $\chi_4$ is much bigger in the active system. The trends in the timescale and the magnitude of heterogeneity are tracked through the peak time (Fig.\ref{fig:x4_combined}b) and area under the curve (Fig.\ref{fig:x4_combined}c). 
 
The timescale of the peak in $\chi_4$ for both the active and passive mixtures track $\tau_\alpha$ (perhaps falling short of $\tau_\alpha$ as approaching glassy regime), but the magnitude of the heterogeneity is always higher in the active system, as has previously been seen in simulations \cite{Paul2023PNAS} of frictionless particles.
In a given mixture heterogeneity among large and small particles track each other at the noise level in active systems, which is not the case in passive systems (Supplementary Material \cite{Supplementary} Fig.\ref{fig:x4_comp}). 

\begin{figure}[tb]
    \centering
    \includegraphics[width=1.0\linewidth]{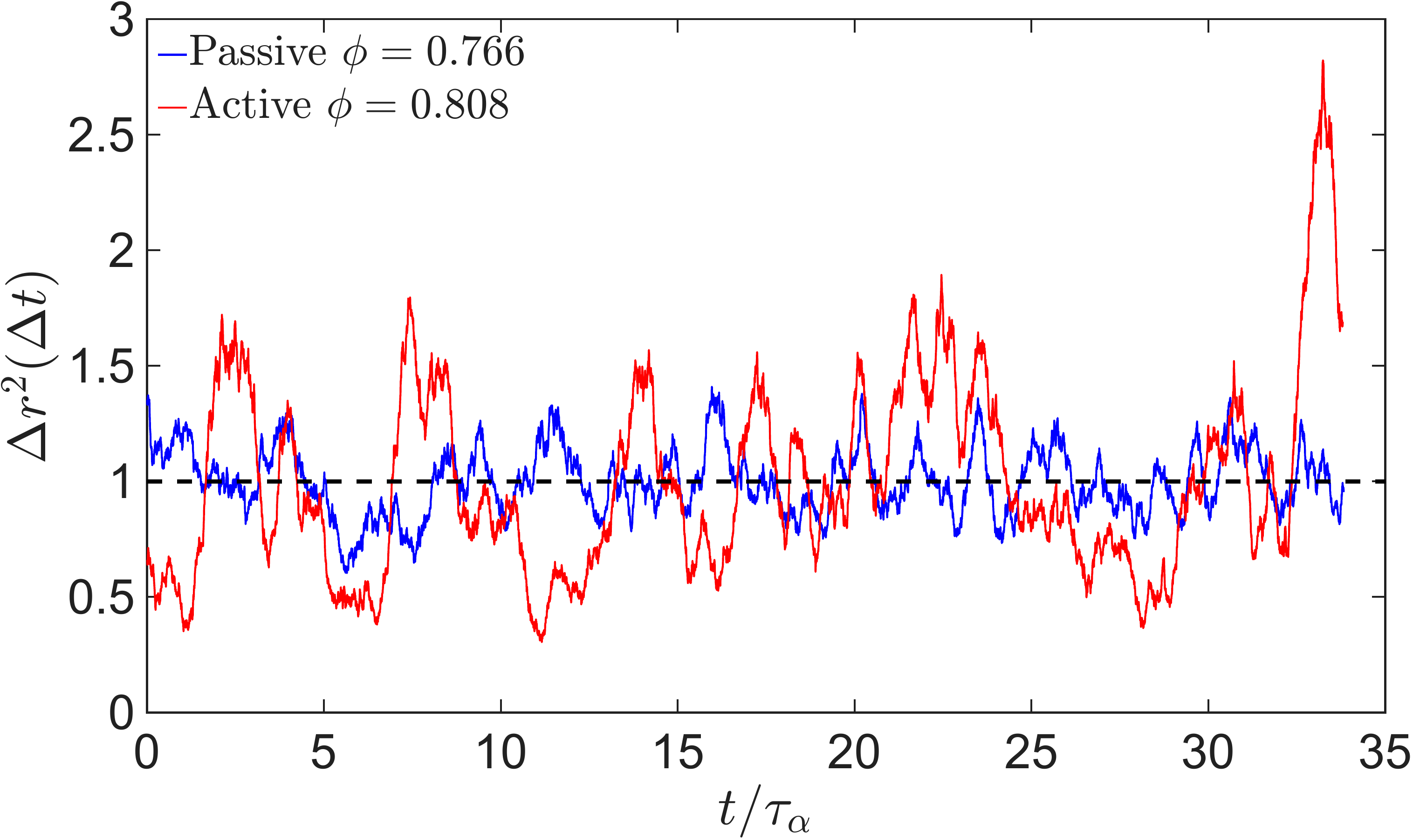}
    \caption{Temporal fluctuations of $\Delta r^{2}(t)$, the squared displacement (normalized over time) computed at a time separation $\Delta t$ = timescale of the peak in $\chi_4$,  averaged over all particles at any instant $t$ (Eqn. \ref{eq:disp_sq}).
    The  temporal fluctuations in the active mixture have larger amplitude.} 
    \label{fig:dis2_all}
\end{figure}

A simpler, and perhaps more intuitive sense of the bigger temporal fluctuations in the active system can be seen (Fig.\ref{fig:dis2_all}) in the time-dependence of the squared displacement, $\Delta r^{2}(t)$, computed at a time interval $\Delta t$ corresponding to the peak timescale of $\chi_4$, when the heterogeneity is the highest. 

\begin{equation}
    \Delta r^2(t) = \frac{\sum_{i=1}^{N} \Delta r_i^2(t)}{\left\langle \sum_{i=1}^{N} \Delta r_i^2(t) \right\rangle_t}
    \label{eq:disp_sq}
\end{equation}

The timescale of fluctuation is similar in both mixtures, but the magnitude is very different, with larger bursts of activity in the active system.

\subsection{Active mixture has higher velocity correlation}
 The observed dynamical heterogeneity is presumably the result of  cooperative particle movement. To capture spatial correlation in particle displacements, we calculated structure-normalized velocity correlations $\Gamma_\textbf{v}(r, \Delta t) = \frac{g_{\mathbf{vv}}(r)}{g(r)}$, as defined below (eqn.\ref{eq:Gamma}). 
$\Gamma(r,\Delta t)$ depends on the timescale $\Delta t$ over which displacement, and therefore velocity is computed (as discussed in Fig.\ref{fig:velr_330} of the Supplementary Material \cite{Supplementary}). Here we select $\Delta t=\tau_\theta$ , the orientational timescale.

\begin{equation}
\Gamma_\textbf{v}(r, \Delta t) 
=
\frac{
\left\langle
\sum\limits_{i,j\neq i}
\mathbf{v}_i(t) \cdot \mathbf{v}_j(t) \,
\delta\!\left(r - r_{ij}(t)\right)
\right\rangle_t
}{
\left\langle
\sum\limits_{i,j\neq i}
\delta\!\left(r - r_{ij}(t)\right)
\right\rangle_t
\left\langle v(t) \right\rangle^2_{t,i}
}
\label{eq:Gamma}
\end{equation}

\begin{figure}[tb]
    \centering
    \includegraphics[width=1.0\linewidth]{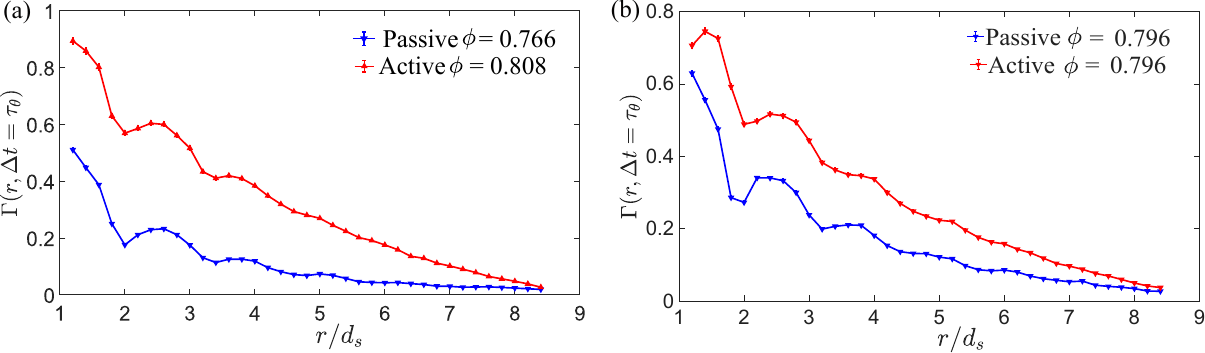}
    \caption{Velocity-velocity spatial correlation, $\Gamma_{\textbf{v}}(r)$, normalized by the radial distribution function, averaged over both small and large particles. The active system shows higher spatial correlations both when compared (a) at the same relaxation time, $\tau_\alpha$, and (b) at same area fraction, $\phi$.}
    \label{fig:velr_pair}
\end{figure}

Compared at the same relaxation time, we find that the active mixture has higher spatial velocity correlations than its passive counterpart (Fig.\ref{fig:velr_pair}). This result is in line with findings of previous simulations \cite{Berthier2019JCP},\cite{Paul2023PNAS}. 

\begin{figure}[tb]
    \centering
    \includegraphics[width=1.0\linewidth]{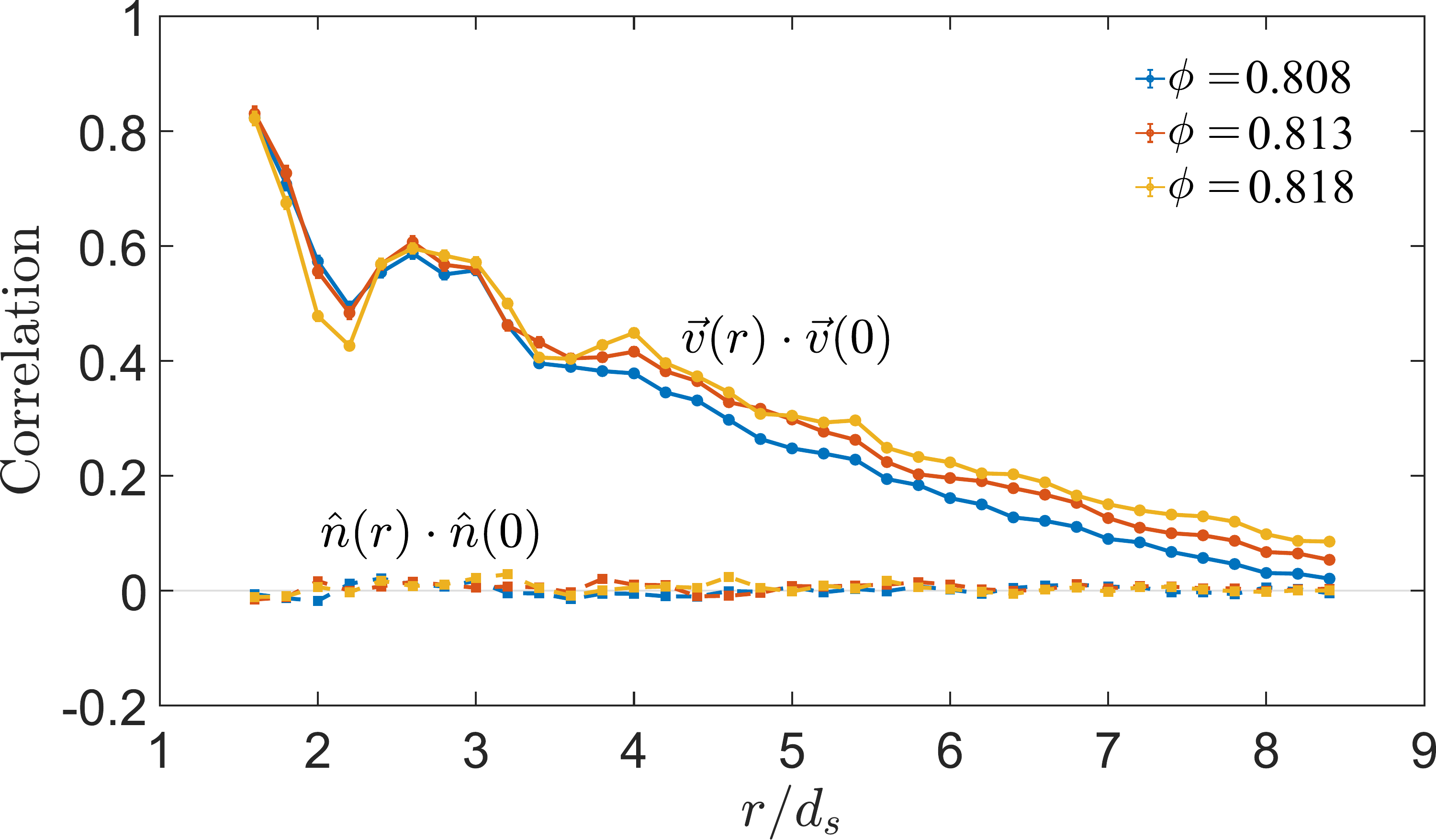}
  \caption{Spatial correlations among large particles in the active mixture at large area fraction.
  Solid lines: Velocity-velocity correlation, $\Gamma_{\mathbf{v}}(r)$, increases modestly with area fraction. Dashed lines: activity-activity correlation $\Gamma_{\hat{n}}(r)$ shows  no alignment in the activity direction at these $\phi$, area fractions.  
}
    \label{fig:velr_lasp3}
\end{figure}

In Figure \ref{fig:velr_lasp3}, we plot the velocity correlation for a few area fractions in the glassy regime: the range of correlations grows modestly with $\phi$ (Velocity correlations $\Gamma_\mathbf{|v|}$ and $\Gamma_\mathbf{\hat{v}}$ also increases with area fraction, Supplementary Material \cite{Supplementary} Fig.\ref{fig:corr_all_lasp_pass}
). In addition to $\Gamma_\textbf{v}$, the spatial correlation of velocity $\mathbf{v}$, we also computed $\Gamma_\mathbf{|v|}$ and $\Gamma_\mathbf{\hat{v}}$ the spatial correlations of  velocity magnitude $\mathbf{|v|}$ and the velocity direction $\mathbf{\hat{v}}$, respectively. We find that all three have comparable length scales of decay  in the glassy regime, with $\Gamma_\mathbf{|v|}$ being modestly larger (see supplementary \cite{Supplementary}-Fig.\ref{fig:corr_325_both}). (This result differs from  recent studies of crowd dynamics \cite{Gu2025Nature} where they observe that velocity directions are spatially correlated over a longer range than speeds. Though there are number of differences between the two systems, in terms of the confinement and the unit 'particles', including the complication of human behavior).
 
When we compute correlations of the activity direction $\mathbf{\hat{n}}$, we find nearly zero correlations at all area fractions, as depicted in Figure \ref{fig:velr_lasp3}. Thus, surprisingly, the \textit{ qualitatively different dynamics in the active mixture do not stem from spatial organization of the activity direction}. The mere presence of persistent active particles is enough to induce not just a speed-up of the dynamics but also behaviour that is qualitatively different from its passive counterpart with the same relaxation time. 

\subsection{Displacement is not aligned with activity}
\begin{figure}[tb]
    \centering
    \includegraphics[width=1.0\linewidth]{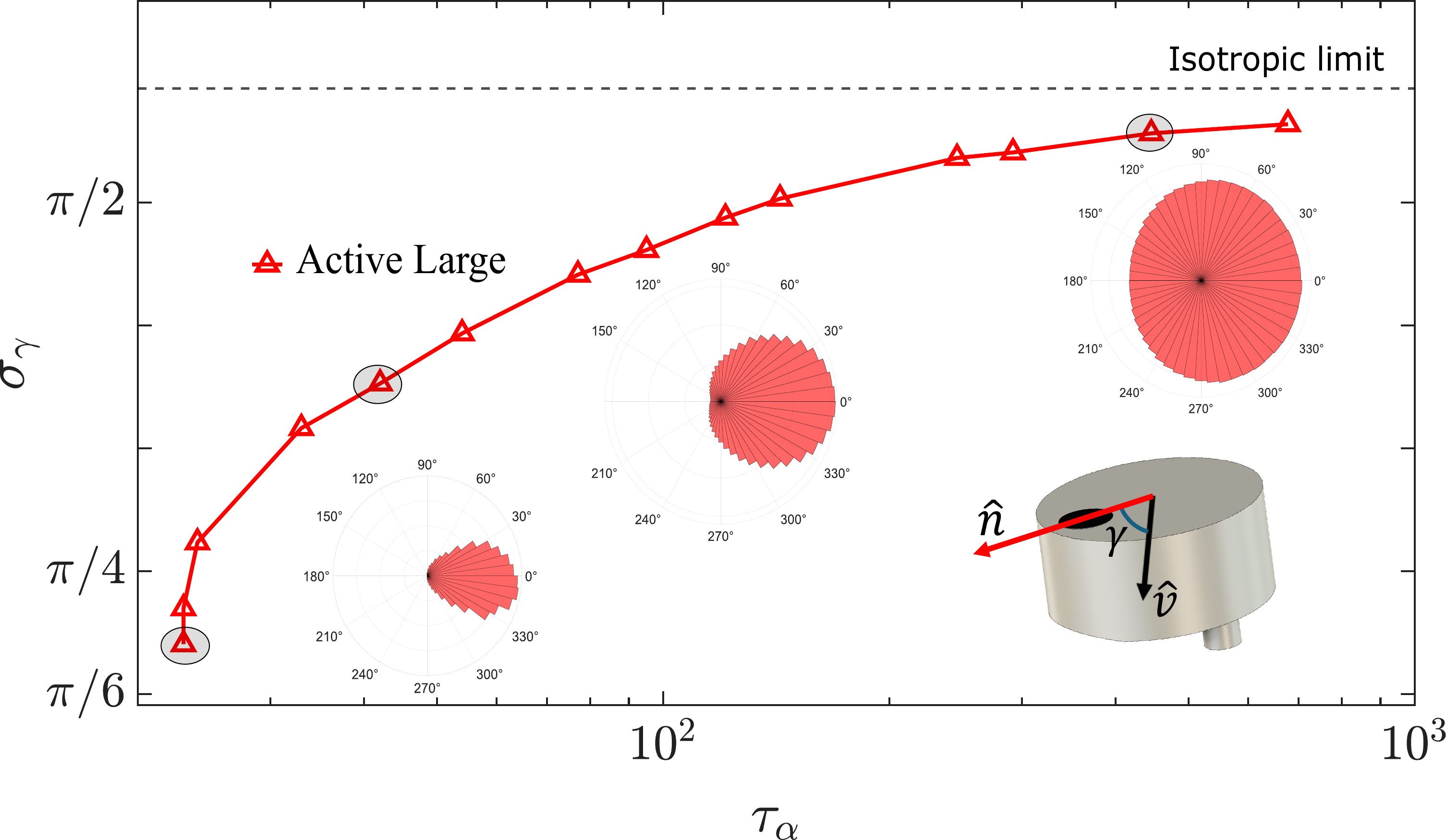}
    \caption{The three polar plots show the distribution of the angle $\gamma$ between the activity direction ($\hat{n}$) and the displacement ($\hat{v}(t)$) in active mixtures with small, intermediate and large relaxation times (indicated by gray circles), where $\hat{v}$ is calculated at a time interval $\Delta t = \tau_\theta^0$, the rotational relaxation time of an isolated active particle.
    The y-axis shows the standard  deviation $\sigma_\gamma$ of the angle $\gamma$ as a function of the relaxation time $\tau_\alpha$ . Both the standard deviation and the polar plots show that in the glassy regime, the correlation diminishes to the isotropic value where the particle displacement is indifferent to the orientation $\hat{n}$.}
    \label{fig:act_vel_ang}
\end{figure}

We now explore the alignment between the activity direction and the local displacement direction at the level of a single particle. By design, in the dilute limit, the displacement direction $\hat{v}(t)$ of an active disc is preferentially along its activity direction $\hat{n}$, as shown in the polar plots in Fig.\ref{fig:exp}(b).  When the angle $\gamma$ between these two directions is computed as a function of area fraction $\phi$, the distribution  $P(\gamma)$ becomes more and more isotropic as depicted in the polar plots embedded in Fig. \ref{fig:act_vel_ang}. (The displacements here are measured over the time scale $\Delta t = \tau_\theta^0$)The standard deviation of the angle distribution approaches the isotropic limit, $\pi/\sqrt{3}$  (Fig.\ref{fig:act_vel_ang}), namely, \textit{in the glassy regime, the displacements of active particles are no more aligned with the active direction}. 

\subsection{Slowing rotational dynamics}
A trivial route by which a particle can lose polar activity is to have a shorter reorientation time, $\tau_{\theta}$. 
However, the loss of polar motions in the translation dynamics is not because of a diminishing persistence time. On the contrary, the persistence of the activity direction increases with the area fraction $\phi$. 
In simulations, the persistence time is held fixed as a control parameter, independent of area fraction, $\phi$. However, in our experiments, the activity direction of a disc is affected by interparticle collisions due to frictional interactions; thus, $\tau_{\theta}$ can change with the area fraction. 
In the glassy regime, particles are caged with a local environment that sustains over the  timescale $\tau_\alpha$ associated with translational relaxation: collisions within this environment, or during the rearrangements associated with cage breakup can affect the activity orientation. 

 \begin{figure}[h!]
    \centering
    \includegraphics[width=1.0\linewidth]{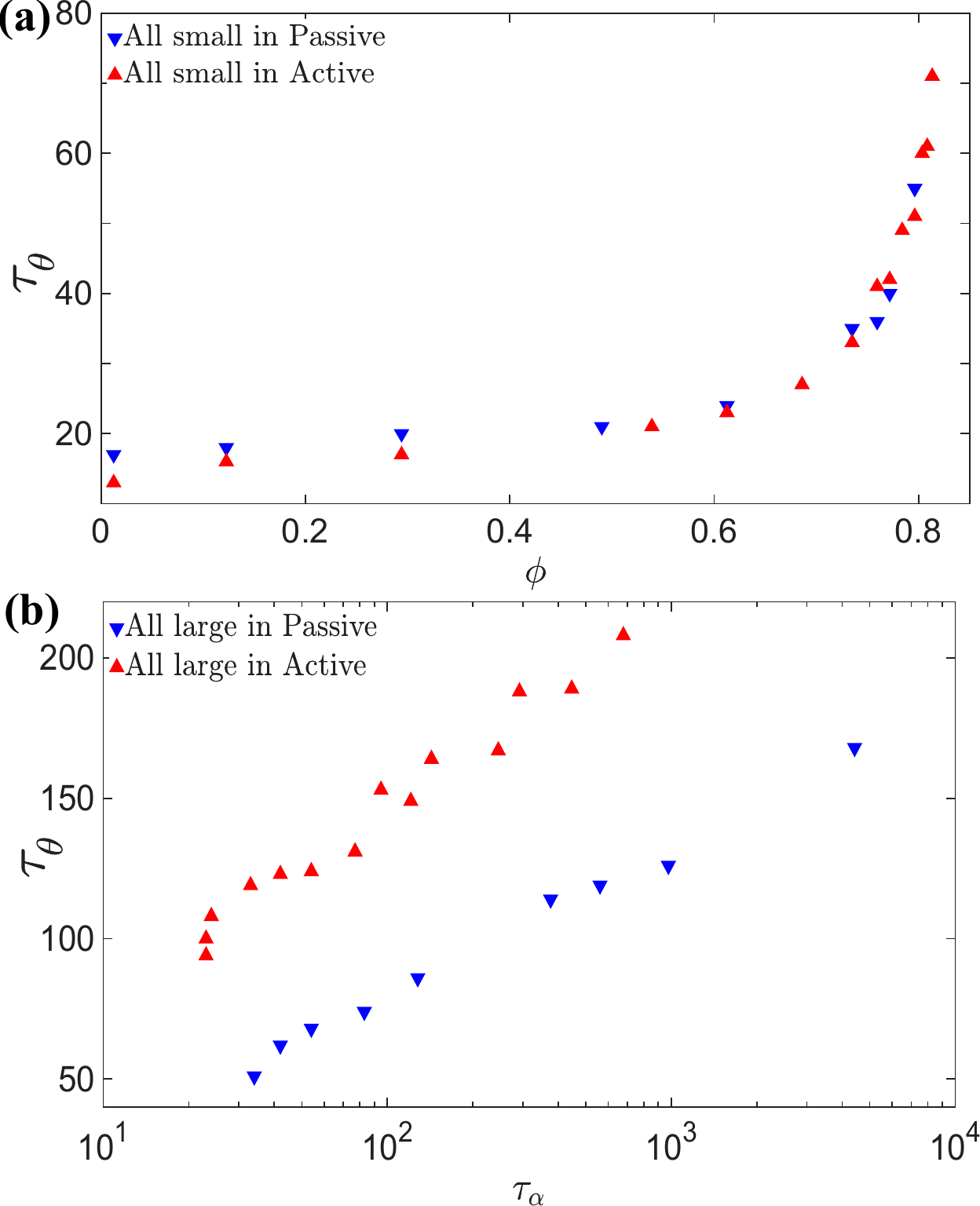}
    \caption{(a) The rotational relaxation time $\tau_\theta$ of the small particles in the two mixtures is very similar at the same area fraction $\phi$ even though their translational relaxation differs. This suggests that rotation is controlled by the local structure, which is similar in both systems.
    (b) In comparing the large particles in the two mixtures, however, at the same relaxation time $\tau_\alpha$, the active large particles rotate more slowly than passive large particles, as per their design. Importantly, despite repeated collisions the persistence of the activity orientation increases, rather than decreases in the glassy regime. 
    \label{fig:msr_all_combine}}
\end{figure}

We find that in both the active and the passive mixture, the  orientational relaxation dynamics slow down for both particles types, whether active or passive.  In Fig.\ref{fig:msr_all_combine}, we show the slowing down of the orientational relaxation time $\tau_\theta$ as the glassy regime is approached: Fig.\ref{fig:msr_all_combine}(a) displays this result for the small particles in both mixtures as a function of $\phi$, and Fig.\ref{fig:msr_all_combine}(b) shows the corresponding results for the large particles as a function of $\tau_\alpha$. 
The orientational dynamics only slow down by a factor of three as compared to a much greater  slowing down in the translation dynamics, as can be seen by comparing the axes. Thus the frictional coupling of rotation and translation is much weaker than in glass formers that follow a Stokes-Einstein-Debye relation. (See Supplementary material \cite{Supplementary} Fig. \ref{fig:msr_toplow_combined},\ref{fig:C_tran_rot_l} for rotation-translation coupling at a given relaxation time)

The orientational dynamics of small particles in the active and passive mixtures are comparable at the same $\phi$ (Fig.\ref{fig:msr_all_combine}(a)) even though the translation relaxation is different (Fig.\ref{fig:rlx} ), suggesting that the rotation are determined by local spatial local structure which is the same in the two mixtures.

\section{Discussion}
We have compared the approach to glassy regime in passive and active mixtures. In our experiments we operate at relatively low activities with a single-particle persistence length of $l_p \sim 4 d_s$. 

Even at these modest activities the relaxation dynamics are much sped up in the active system. But when passive and active mixtures with the same relaxation times are compared, the active system is qualitatively different from the passive system, with different relaxation shape, substantially larger dynamical heterogeneities, and longer range velocity correlations.  Remarkably, though, this happens with no organization of the activity direction of the active species. Indeed, the defining feature of an active particle -- that it move along the activity axis -- is completely suppressed in the crowded environment of the glass regime. 

Thus the picture that emerges of the large collective motions observed in the active glass is not one in which motion is preferentially driven along the activity axis, but by persistent anisotropic collisional stresses provided by the active particles even though these are globally unpolarized.  Thus persistent local pushes lead to avalanches in motion which are not at all like the classic active dynamics of MIPS or flocking which are closely locked to activity orientation.  Rather, the dynamics we observe are much more reminiscent of those in confluent epithelial tissues \cite{bi2016motility} or in dynamics of  crowds \cite{Gu2025Nature} where large swirls of activity are observed even though the  bodies in the crowds show no obvious alignment.  Recently, some aspects of the collective dynamics in both passive \cite{weeks2017mermin} and active systems \cite {Karmakar2025marmin} have been ascribed to the long-wavelength fluctuations that destabilize low-dimensional ordered systems \cite{merminWagner1966}. These Mermin-Wagner fluctuations also can appear in disordered systems. A local measure that separates out long-wavelength modes is to compute the mean-squared displacement $\Delta r^{2}(t)$ relative to the local neighbourhood \cite{Karmakar2025marmin} ; these cage-relative MSD's 
(Supplementary Material \cite{Supplementary} Fig.\ref{fig:CRMSD}) are lower than the conventional MSD (almost by definition). Whether this uniquely points to Mermin-Wagner fluctuations is an issue that merits further exploration.  

 A couple of features  are unique to the granular context: one is the role of inelasticity, which in less dense situations has been known to produce clumping and velocity correlations \cite{olafsen1998clustering}; inelastic clumping has also been revisited in the context of the glass problem\cite{Komatsu2015PRX}. However, inelastic dissipation is common to both the active and passive mixtures, and their dynamics are therefore  distinguished by the presence or absence of activity. A second distinctive feature of frictional particles is that the orientational degree of freedom is coupled to collisions, and therefore the orientational relaxation time changes with area fraction $\phi$.  A crucial factor here is that despite collisional coupling to the activity direction, the relaxation time increases (rather than decreasing with collisions).  The persistence of the activity direction in providing sustained anisotropic kicks is much more consequential for the active glass than the short persistence length of the particles, which becomes irrelevant in the crowded, caged dynamics of the particles. Thus the picture that emerges is that activity generates sustained, uncoordinated, local, anisotropic stresses - the highly-coupled physics of this dense medium does all the rest. 

\appendix
\section{Experimental Setup}
We provide further experimental details in this section.\\
\textbf{Cell and particles}: The nylon mesh glued to the base has an opening size of $0.29$ mm and wire diameter of $0.21$ mm, which is finer than the dimensions of the pins on the discs, which are of diameter = 1.27 mm. The pin-to-body height ratio is, Large Passive $\sim$ 0.49, Large active $\sim$ 0.26 and in Small passive $\sim$ 0.31. In addition to suppressing slippage of discs against the floor,  the mesh also has the effect of making the short time displacement fluctuations to be gaussian Fig. \ref{fig:exp}(c). \\
The lid that confines the particles is transparent, static-dissipative cast acrylic that minimizes tribocharging. To maintain optical clarity, we use an abrasion resistant and UV resistant grade of acrylic. \\
The scalloped boundary that suppresses ordering and slip at the boundary is made of acrylic of thickness $3.125$mm, has an outer diameter of radius $20$cm, inner geometry scalloped with local radius of curvature $9$mm. The radius
of curvature is small enough such that there is no flow due to
recirculation at the boundary.\\ 
The particles are 3D printed using EOS P110, in a powder bed nylon printer that uses selective laser sintering (SLS). This printing technique does not require support material, and has a resolution of $\approx 100 \mu m$ so that the details of the pin and the edges of the disc are reproducibly rendered. These are essential for our experiments as collisions of the discs can be significantly altered by excess material or poor definition of surface features\\

\textbf{Image acquisition and analysis}
We record the video at 125fps using Photron FASTCAM mini UX100 camera positioned for a top view of the cell. 
Video frames are saved as uncompressed grayscale TIF files. 
We use custom MATLAB code for image analysis. We first process the image with this sequence of operations: \\
 Crop $>$ Normalize the grayscale intensity $>$ Scale by a factor $>$ Binarize by setting a threshold $>$ Use MATLAB inbuilt imfindcircles function to detect the circles.\\
We used a  similar process to detect the small dots on the discs. Once we have the centroids of the circles at every frame we build the trajectories of every particles by setting a maximum search distance which we set to be $0.3d_s$ so there is no mismatch. The dots are matched with the particle by setting the search distance to be around the actual distance between the centroid of the particle and the dot position, which is fixed and uniform among particles. 

\textbf{Granular temperature}:
The short time dynamics of the active and passive discs are nearly gaussian with zero mean and the distribution is very similar for the two cases.
Table \ref{tab:displacements} shows the standard deviation of displacement and rotation after one shake. 

\begin{table}[h]
\centering
\caption{Displacement and rotation after one shake 
}
\label{tab:displacements}
\begin{tabular}{lcc}
\hline
\hline
Quantity & Large Passive & Large Active \\
\hline
$\sigma(\Delta\theta)$ (rad) & $0.039$ & $ 0.041$ \\
$\sigma(\Delta r_x)$ (ds) & $0.038$ & $0.041$ \\
$\sigma(\Delta r_y)$ (ds) & $0.065$ & $0.073$ \\
\hline
\hline
\end{tabular}
\end{table}

\section{Finite-size effects}
\label{app:area_fraction}

\begin{figure}[h]
    \centering
    \includegraphics[width=1.0\linewidth]{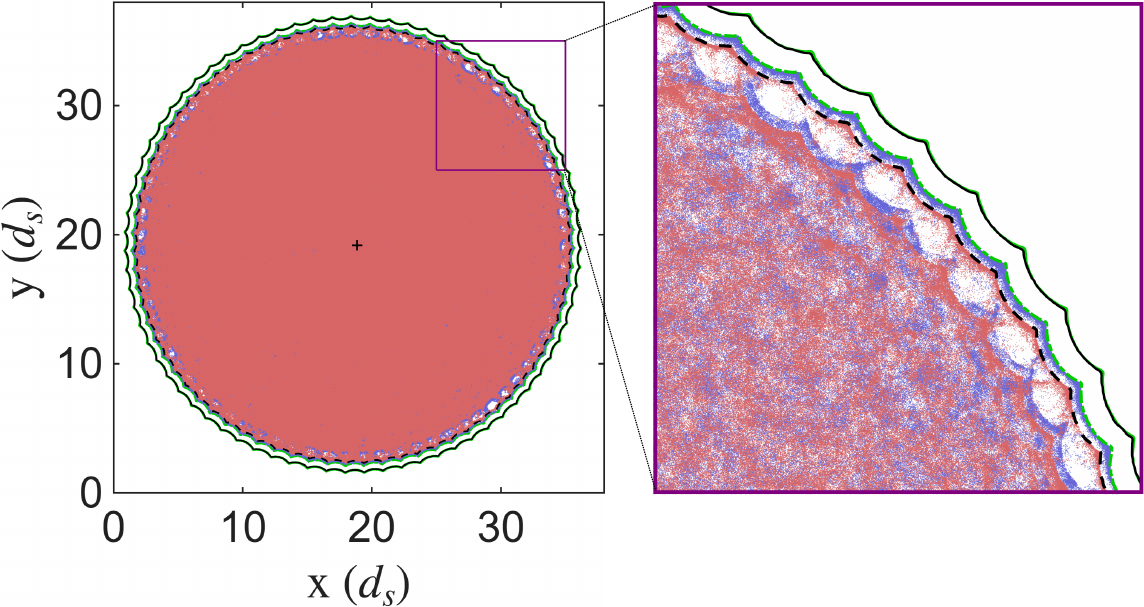}
    \caption{The left panel shows a scatter plot of the centroids of the small and large particles at $\phi = 0.612$, an area fractions at which particles explore  the entire accessible area of the cell. The right panel zooms in to show particle centroids in the vicinity of the scalloped boundary. We detect the outer envelope of occupied position for both large  and small particles (indicated by black and green dashed lines for large and small particles respectively). We approximate the boundary of the accessible area by adding the radius of the corresponding species (solid black and green lines). The enclosed areas, $A_s$ and $A_l$, are used to compute area fraction $\phi$.}\label{fig:boundary}
\end{figure}

\textbf{Calculation of area fraction $\phi$}\\
The boundary of our system is bumpy, as shown in Fig \ref{fig:exp}. To calculate the area fraction, we need to determine the area accessible to particles. We choose a large volume fraction, where particles push against the boundary and find particle centroid positions over time (Fig.\ref{fig:boundary}, left panel). We then fit the outer envelope of the centroid positions to get the shape accessible to particle centres and add the corresponding particle radii to obtain the total areas, $A_l$ and  $A_s$, accessible to the  small and large discs, respectively (Fig.\ref{fig:boundary} right panel).

We determine the mean projected areas $a_s$ and $a_l$ of small and large discs averaged from images in the dilute regime. These projected areas are not identical to the area of the circular part of the discs as the quasi-2D confinement allows the discs to be tipped when in motion. 
\begin{equation}
    \phi = N\left(\frac{a_s}{A_s}+\frac{a_l}{A_l}\right) \ \ , 
\end{equation}
where N is the number of particles of each size. 

\textbf{Boundary effects on the packing}
As discussed earlier Desmond et al. \cite{Desmond2009PRE} developed a semi-empirical model which quantifies the reduction in packing due to finite size effects, parametrized for  circular confinement by the ratio $h$ of system diameter to small particle diameter. For our system $h = \sqrt{\frac{A_s}{a_s}} = 31$. For this system size, the random packing area fraction is  $\phi(h=31) = 0.823$ in comparison to $\phi_{rcp} = \phi(h \rightarrow \infty) = 0.846$.

\begin{figure}[h]
    \centering
    \includegraphics[width=1.0\linewidth]{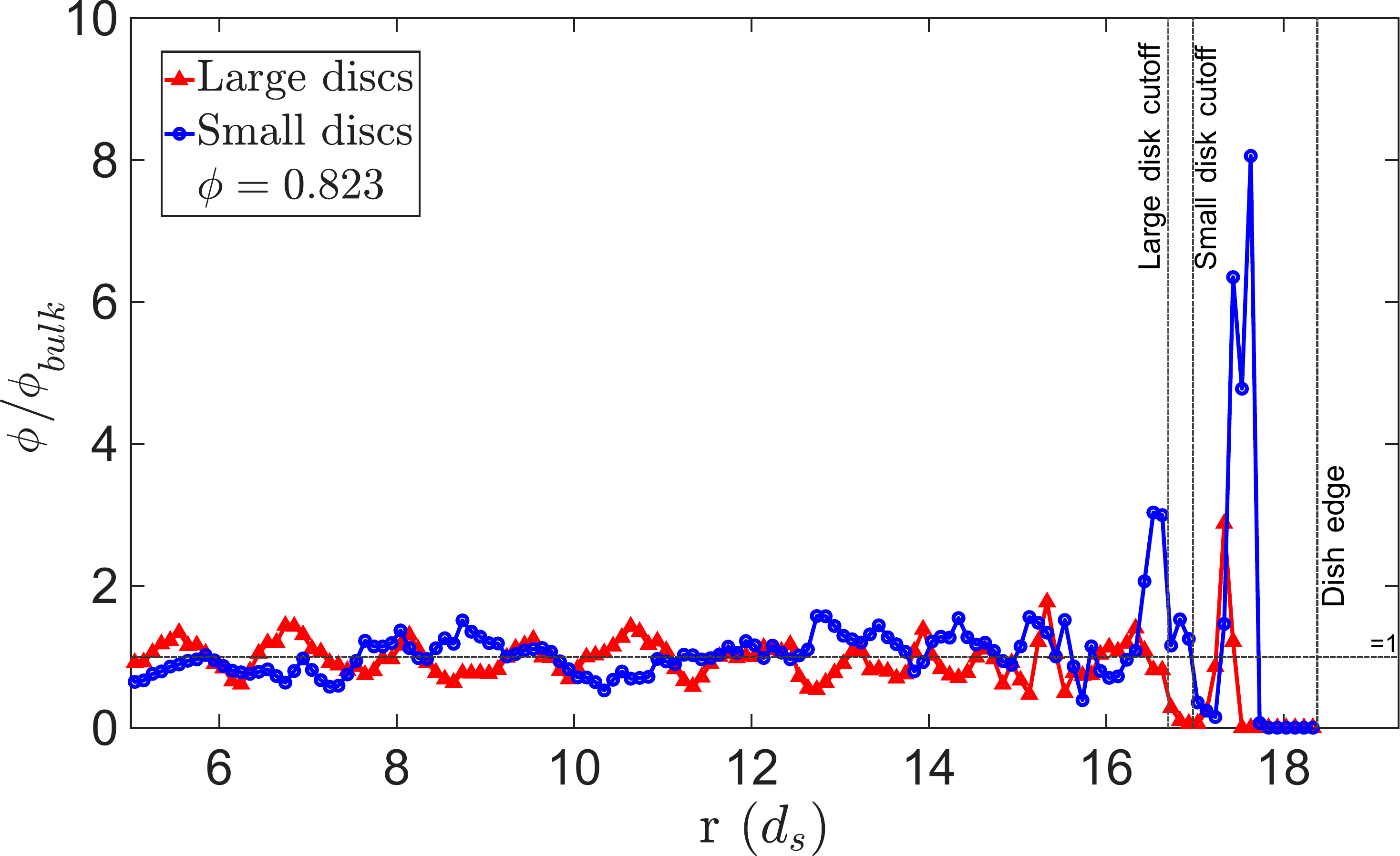}
    \caption{The wall-induced structure is quantified by the radial dependence of the normalized area fraction $\phi$. The normalized fraction fluctuates around 1 in the bulk, but peaks near the boundary. In all the analysis we have excluded this outer high-density layer in the vicinity of the boundary.} 
    \label{fig:weeks_rcp}
\end{figure}

In Fig. \ref{fig:weeks_rcp}, we show the radial dependence of density for both small and large particles. As noted earlier, there is wall-induced structure which leads to a peak in density for both the particles at the boundary. In all the analysis, we remove particles in this peak at the boundary as both their structure and dynamics are not typical of the bulk. 

\section{Rotation}
\label{app:rotation}
We can characterize rotational dynamics by two methods.  One conventional method is to compute the autocorrelation of the orientation vector $\left\langle \hat{n}(t)\cdot\hat{n}(0)\right\rangle$. For a diffusive, memoryless noise, the autocorrelation declines exponentially in time with a timescale $D_R^{-1}$ (see e.g. \cite{Walsh2017SM}.

\begin{figure}
    \centering
    \includegraphics[width=1.0\linewidth]{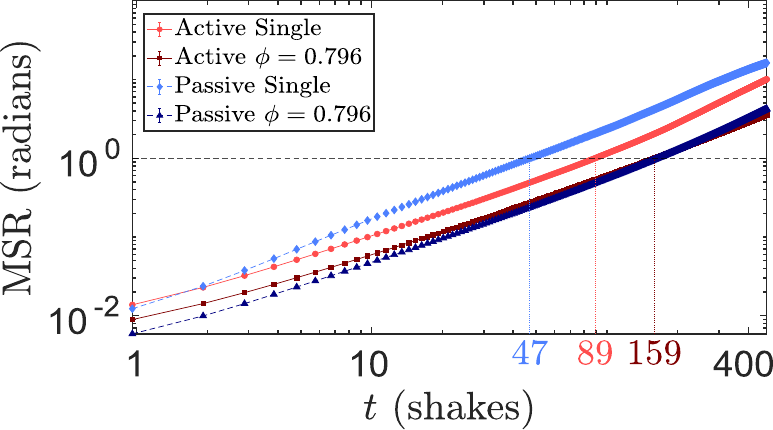}
    \caption{The mean square rotation(MSR) $ \left\langle\Delta\theta^2(t)\right\rangle$ is determined from the orientation $\theta(t)$ of the direction $\hat{n}$ in the lab frame.  Here $\theta(t)$ is the cumulative rotation angle and is not wrapped to $[0,2\pi]$. The orientational relaxation timescale $\tau_\theta$, is defined as the time when $MSR, \left\langle\Delta\theta^2(t)\right\rangle$=1. As examples, we show MSR for large active and passive particles in the dilute limit and averaged over all large particles at area fraction $\phi = 0.796$ in the both mixtures.}
    \label{fig:MSR_DR}
\end{figure}

An alternative method, which we adopt in this paper, is to compute the mean square rotation $MSR, \left\langle\Delta\theta^2(t)\right\rangle = \left\langle \frac{\sum_{i=1}^{N}  (\theta_i(t)-\theta_i(0))^2}{N} \right\rangle$, where N is the number of particles and the average $\left\langle\right\rangle$ is taken over the initial time $0$, and particles. We then define the rotation timescale $\tau_\theta$ as the delay time when $\left\langle\Delta\theta^2(t_{\theta})\right\rangle = 1$ as shown in Fig. \ref{fig:MSR_DR}. A feature of the MSR is that we can unwrap the total rotation angle of particles, rather than maintain them only modulo $2\pi$ as in the autocorrelation of $\hat{n}(t)$. 

 We already saw in Fig. \ref{fig:msr_all_combine} that crowding slows down rotational relaxation. In Fig. \ref{fig:MSR_DR}, an explicit comparison is shown of  $\tau_\theta$ via the single-particle and finite-density MSR.  Fig. \ref{fig:MSR_DR} also shows that short-time limit of the MSR is the same for the isolated active and passive particle, as shown earlier in Fig. \ref{fig:exp} (c). However, the MSR for the active particle has slower time dependence than the passive particle because of the persistent motion. The microscopic mechanism for  persistence could possibly be self-alignment \cite{Baconnier2025RMP}.

\section{Persistence Length}
\begin{figure}[h!]
    \centering
    \includegraphics[width=1.0\linewidth]{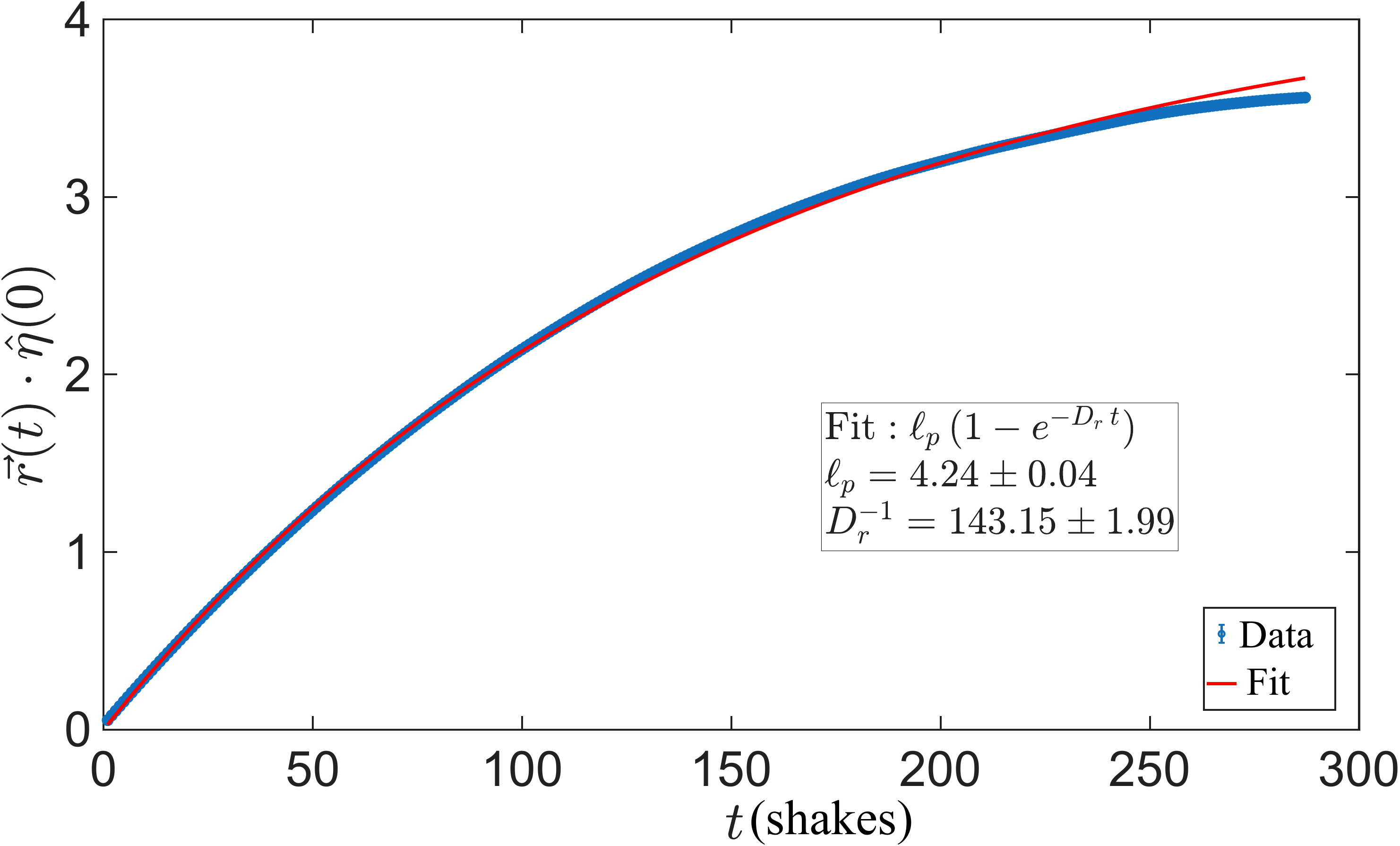}
    \caption{Orientation-displacement correlation of an isolated active particle. We fit these data to an exponential form $l_p(1-e^{-D_r t})$ determine the persistence length $l_p$, which is $\approx 4 d_s$ for the active particles used here.}
    \label{fig:OD_corr}
\end{figure}

We obtain the persistence length $l_p$ of the active particle by computing the orientation-displacement correlation$\left\langle \vec{r}(t)\cdot\hat{n}(0)\right\rangle$ in the dilute limit  (Fig.\ref{fig:OD_corr}). We extract $l_p$ by fitting to $l_p(1-e^{-D_rt})$, obtaining a persistence of length $ \approx 4.2d_s$. We have used $D_r$ as a free parameter in this fit, rather than extracting it from the autocorrelation of $\hat{n}(t)$, due to small deviations from the exponential function expected for delta-correlated noise \cite{Walsh2017SM}. 

\begin{acknowledgments}
We gratefully acknowledge support from NSF-DMR 2319881.  We thank C.S. O'Hern and K. Ramola for questions that led to the current versions of Fig.\ref{fig:exp}C and Fig.\ref{fig:velr_lasp3}, respectively. 

\end{acknowledgments}

\bibliography{references}

\clearpage
\newpage

\setcounter{section}{1}
\setcounter{figure}{1}
\setcounter{table}{1}
\setcounter{equation}{1}

\renewcommand{\thesection}{S\arabic{section}}
\renewcommand{\thesubsection}{\thesection.\arabic{subsection}}
\renewcommand{\thefigure}{S\arabic{figure}}
\renewcommand{\thetable}{S\arabic{table}}
\renewcommand{\theequation}{S\arabic{equation}}

\section*{S1 Supplementary Material}
\addcontentsline{toc}{section}{S1: Supplementary Material}

\subsection{Width of relaxation of both species}
\begin{figure}[h!]
    \centering
    \includegraphics[width=0.8\linewidth]{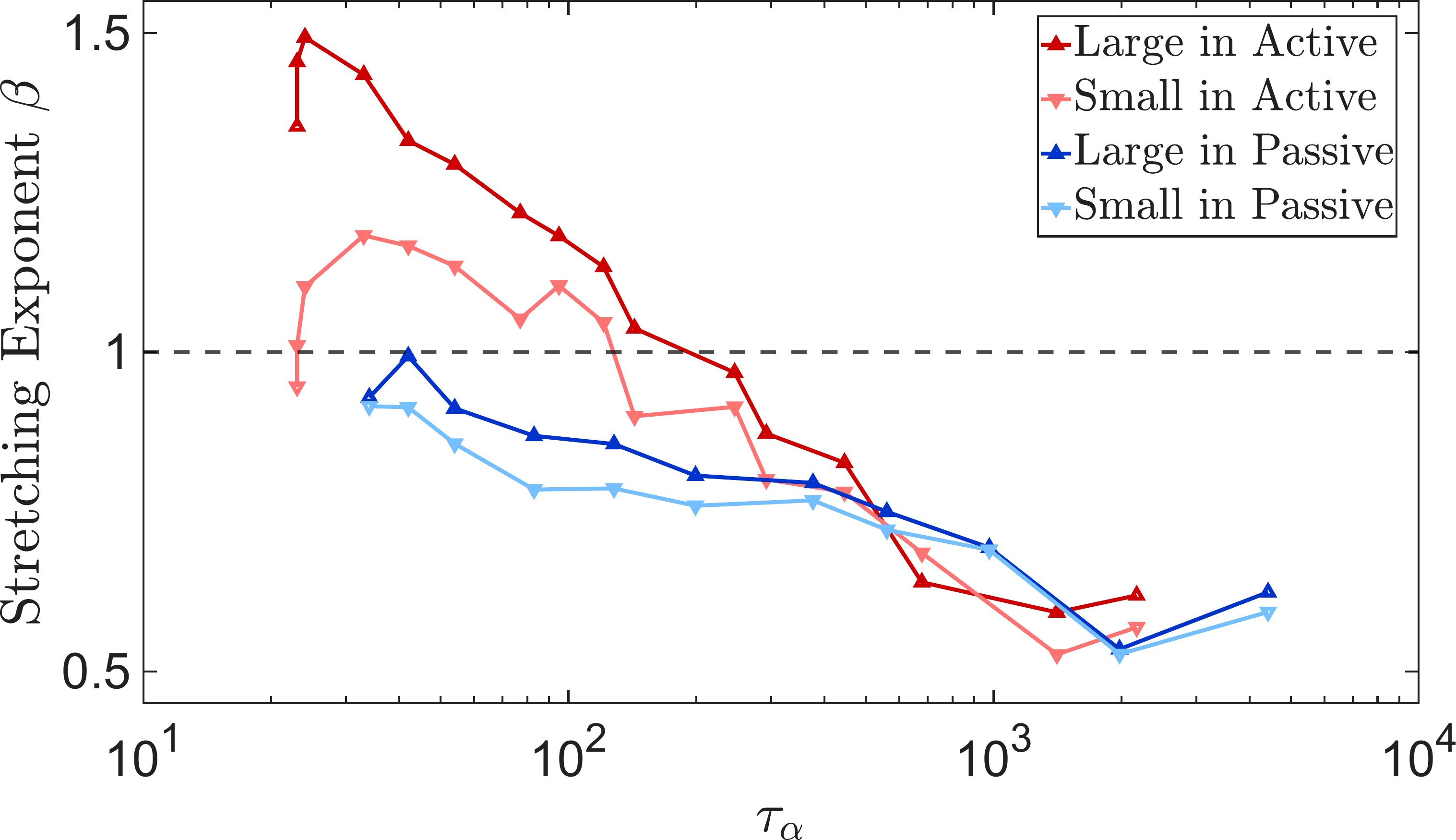}
    \caption{Stretching exponent $\beta$ for large and small species in both mixtures plotted against relaxation time $\tau_\alpha$ of the large discs at that area fraction. }
    \label{fig:beta_both}
\end{figure}

In the main text (Fig.4), we showed results from fitting the shape of the self-intermediate scattering function of only the large particles in the two mixtures,  to the stretched exponential function: $F_S(q,t)=exp(-(t/\tau_{\alpha})^\beta)$. 
In Fig.\ref{fig:beta_both} we show $\beta$ for the small passive particles along with those of large particles. We recall that the decreasing exponent $\beta$ signifies a broadening relaxation. Here we see that the relaxation of the small particles in the dilute limit is nearly exponential. However, as the mixtures get denser, $\beta$ for both components in the mixture track each other.  

\subsection{Rotation-translation coupling}
\begin{figure}[h!]
    \centering
        \includegraphics[width=1.0\linewidth]{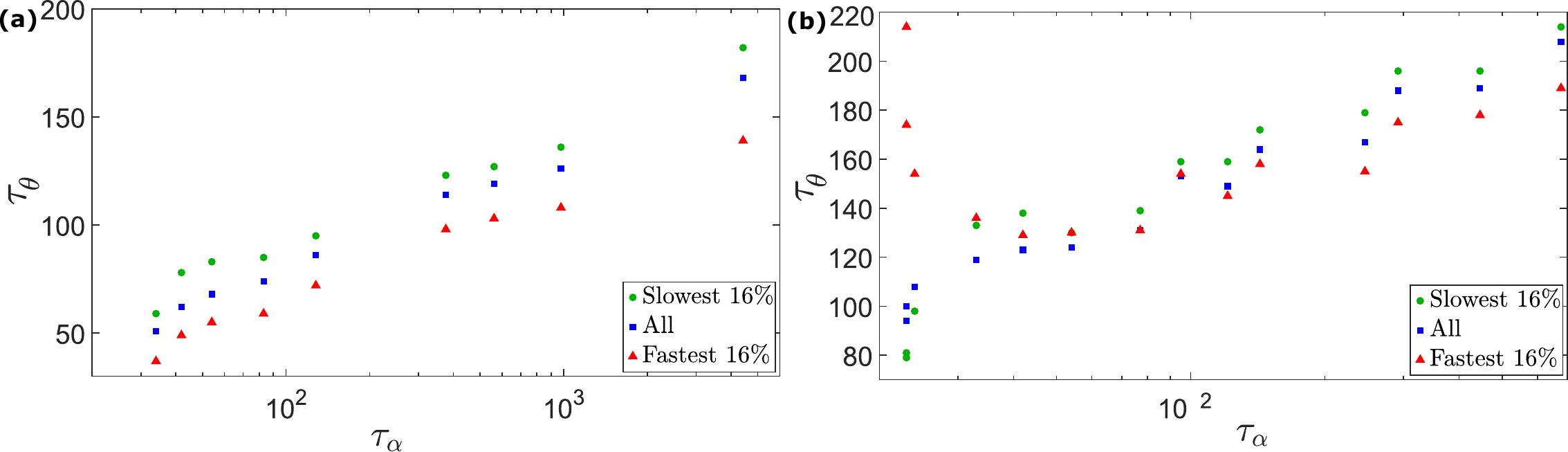}
    \caption{Rotational relaxation time of the fastest, slowest and average particles in the mixture, plotted against $\tau_\alpha$. (a) Large particles in the passive mixture. Fast translating particles also rotate faster. For a given relaxation time, corresponding to a given area fraction, the fastest-translating particles rotate faster than the mean, which in turn is faster than the slowest 16\% large particles.
             (b) Large particles in the active mixture. Crossover of rotational relaxation for fast and slow large \emph{active} particles as a function of relaxation time. In the dilute limit fast moving particles  rotate slower. There is a crossover after an intermediate area fraction, after which the fast translating particles also rotate faster.}
    \label{fig:msr_toplow_combined}
\end{figure}

In this section, we further quantify the rotation-translation coupling by comparing the rotation of the fast-moving particles to the slow particles. We rank the translational mobility of the particles based on the mean square displacement calculated at lag time $\tau_{\theta}$. We compare the average rotation of the top $16\%$ fast moving particles to the rotation of slowest $16\%$ particles. 
In the passive system, we find that fast-moving particles are also rotating faster than slow moving particles as shown in Fig.\ref{fig:msr_toplow_combined}(a) where we plot $\tau_\theta$ against $\tau_\alpha$ separately for fast- and slow-moving particles. 
In the active system, Fig.\ref{fig:msr_toplow_combined}(b), however, there is a crossover: in the dilute regime fast-moving particles persist in direction, i.e. rotate less. However, at intermediate and higher area fractions, fast particles also rotate fast, just as in the passive system. \newline

Rather than segmenting the population of particles in this manner, we can quantify this relationship by calculating a covariance coefficient between rotational  and translational motions:
\begin{equation}
    C = 
        \frac{
            \sum_{i=1}^{n} 
            (\Delta r_i - \overline{\Delta r})
            (\Delta \theta_i - \overline{\Delta \theta})
        }{
            \sqrt{
                [\sum_{i=1}^{n} (\Delta r_i - \overline{\Delta r})^2]
                [\sum_{i=1}^{n} (\Delta \theta_i - \overline{\Delta \theta})^2]
            }
        }
\label{eq:coeff}
\end{equation}

This average sums over all  particles, unlike the previous calculation where we averaged over fastest $16\%$ and slowest $16\%$. 

\begin{figure}[h!]
    \centering
    \includegraphics[width=0.8\linewidth]{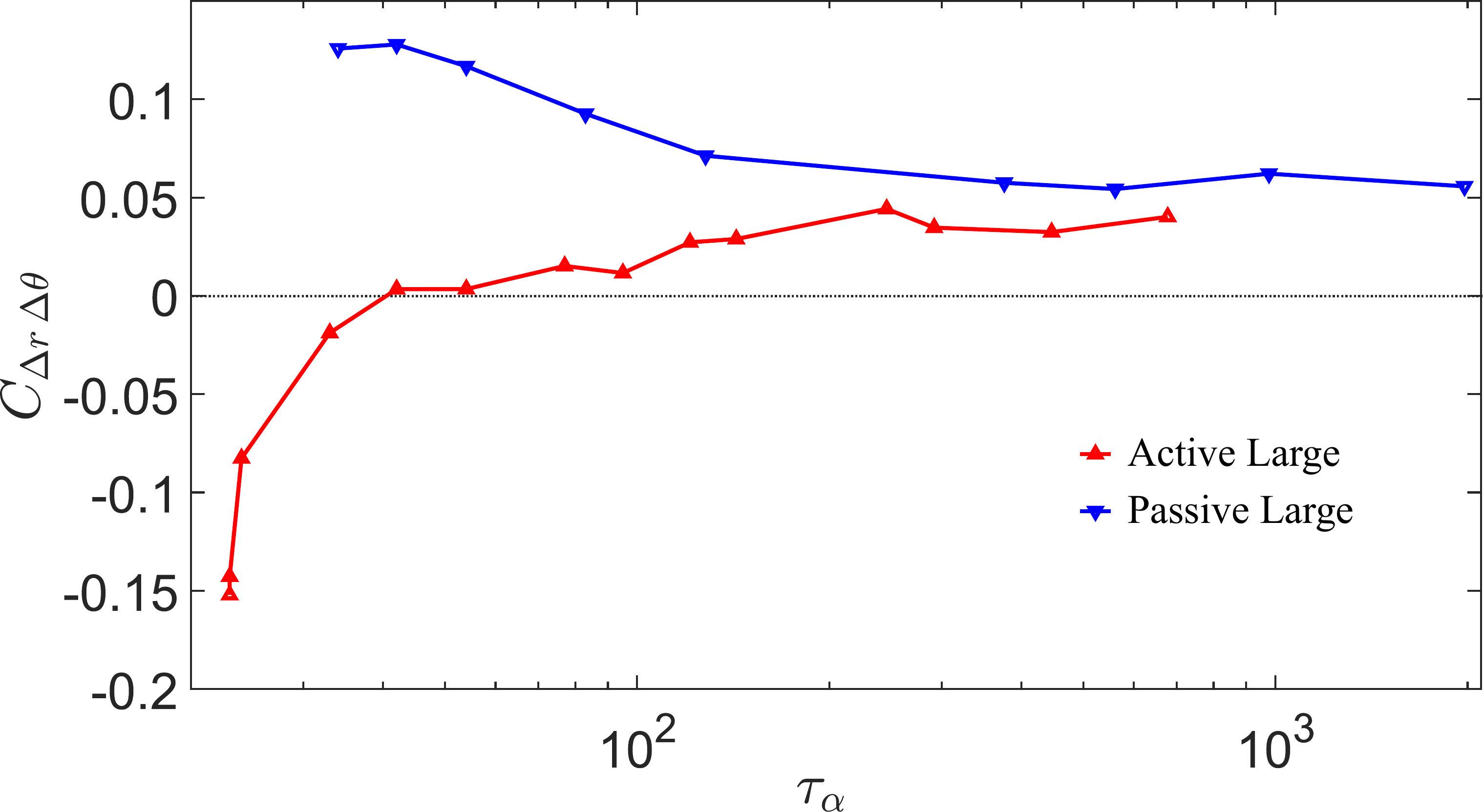}
    \caption{
        Rotation–translation correlation coefficient $C$ (see eqn.\ref{eq:coeff}), where translation displacement $\Delta r$ and rotation $\Delta \theta$ are calculated at the corresponding $\tau_{\theta}$. The correlation coefficient $C > 0$ for the passive system whereas there is crossover in the active system from $C < 0$ in the dilute regime to $C > 0$ at high densities.
    }
    \label{fig:C_tran_rot_l}
\end{figure}

The results are consistent, with the covariance coefficient (Fig.\ref{fig:C_tran_rot_l}) always positive for the passive system, whereas in the active system, the correlation switches from being negative in the dilute regime to positive thereafter. 

\subsection{Cage relative MSD}

\begin{figure}
    \centering
    \includegraphics[width=1\linewidth]{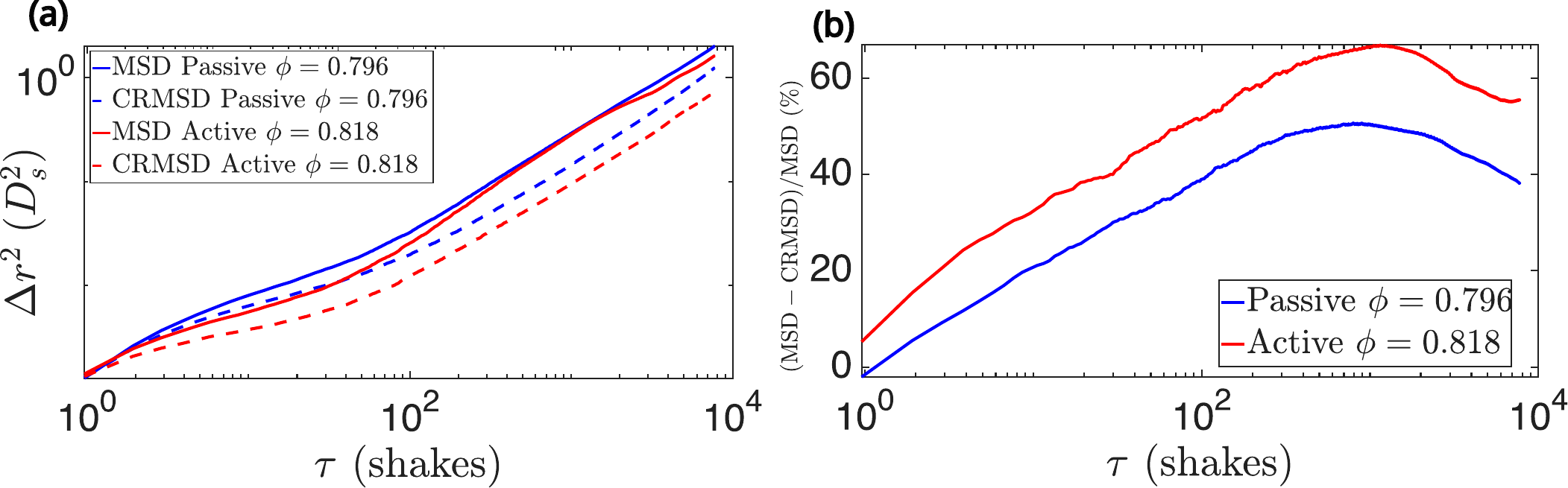}
    \caption{(a)Cage relative MSD (dashed) sits below the MSD (solid) curve for both the systems. This tells us that there is cooperative motion in both the systems. (b) Percentage of cooperative displacement as function of lag time. Cooperative motion goes up with time, saturates around $\tau_{\alpha}$ and then decreases.}
    \label{fig:CRMSD}
\end{figure}

 As discussed in the main text, both passive \cite{weeks2017mermin} and active disordered systems \cite {Karmakar2025marmin} have been shown to host long-wavelength fluctuations that have been ascribed to  Mermin-Wagner fluctuations\cite{merminWagner1966}. In this section, we compute cage relative mean square displacements (CR-MSD) (\ref{eq:cr_msd}) and compare them to the mean square displacement (MSD) (\ref{eq:msd}).  

\begin{equation}
    \Delta r^2(t) = \frac{1}{N} \sum_{i} \left\langle \left| \mathbf{r}_i(t) - \mathbf{r}_i(0) \right|^2 \right\rangle
    \label{eq:msd}
\end{equation}

\begin{equation}
    \Delta_{\mathrm{CR}} r^2(t) = \frac{\sum_{i} \left\langle \left| \mathbf{r}_i(t) - \mathbf{r}_i(0) - \frac{1}{N_i} \sum_{j} (\mathbf{r}_j(t) - \mathbf{r}_j(0)) \right|^2 \right\rangle}{N}
    \label{eq:cr_msd}
\end{equation}
\\
Here, the second sum is taken over the $N_i$ nearest neighbors $j$ of particle $i$, as determined by a Voronoi tessellation \cite{Keta2022PRL}. This has the effect of removing displacements that are also experienced by the particle's immediate neighbourhood.
In Fig.\ref{fig:CRMSD} we plot the MSD and CR-MSD of systems of active mixture ($\phi = 0.818$) and passive mixture ($\phi = 0.796$) which have similar relaxation times. 

We find similar qualitative behaviour in both quantities as the area fraction is varied. Quantitatively, the CR-MSD is lower than MSD. To give a sense of the trends in this term with area fraction, we take the difference of the MSD and CR-MSD divided by the MSD , which gives us the relative percentage of the cooperative displacement. As shown in Fig.\ref{fig:CRMSD}(b), the fractional contribution increases with time, saturates around relaxation time, and then decreases. Our work does not contribute to deciding whether this is a feature of Mermin-Wagner fluctuations, however, the main results of this article are unaltered whether or not we consider this subtraction. 

\subsection{Spatial velocity correlations}

In the main text, we define the structure-normalized  correlations $\Gamma_\textbf{v}(r, \Delta t) = \frac{g_{\mathbf{vv}}(r)}{g(r)}$, as defined below. 
$\Gamma(r,\Delta t)$ depends on the timescale $\Delta t$ over which displacement, and therefore velocity is computed. In this article, we select $\Delta t=\tau_\theta$, the orientational timescale.

\begin{equation}
\Gamma_\textbf{v}(r, \Delta t) 
=
\frac{
\left\langle
\sum\limits_{i,j\neq i}
\mathbf{v}_i(t) \cdot \mathbf{v}_j(t) \,
\delta\!\left(r - r_{ij}(t)\right)
\right\rangle_t
}{
\left\langle
\sum\limits_{i,j\neq i}
\delta\!\left(r - r_{ij}(t)\right)
\right\rangle_t
\left\langle v(t) \right\rangle^2_{t,i}
}
\label{eq:Gamma}
\end{equation}

This definition we presented for $\Gamma$ in the case of velocity correlations readily extends to the correlations of $\hat{n}$ and $\hat{v}$ because they have zero mean, just as the velocity ($\vec{v}$) does. 

For the correlation of velocity magnitude, which has a nonzero mean, we compute and subtract the mean square in the numerator. For a bounded system like ours, there is a further complication, in that the mean speed has a radial dependence over the cell. Taking that into account, we define the velocity magnitude correlation as
\begin{equation}
\Gamma_{|\mathbf{v}|}(r, \Delta t) = 
\frac{
\left\langle
\sum\limits_{i,j\neq i}
\Big[ v_i(t)v_j(t) - \bar v(\rho_i)\bar v(\rho_j) \Big]
\delta\!\left(r - r_{ij}(t)\right)
\right\rangle_t
}{
\left\langle
\sum\limits_{i,j\neq i}
\delta\!\left(r - r_{ij}(t)\right)
\right\rangle_t
\sigma_v^2
}
,
\label{eq:Gamma_mag}
\end{equation}
where $v_i$ is the magnitude of the velocity of the particles $i$, $\bar v(\rho_i)$ is the mean of the speed of the particles at distance $\rho$ from the center of the cell, and $\sigma_v^2$ is the variance of the speed. As in the other correlators (e.g. $\Gamma_\textbf{v}(r, \Delta t)$), here too, we use $\Delta t=\tau_\theta$.\\

\begin{figure}[h!]
    \centering
    \includegraphics[width=1\linewidth]{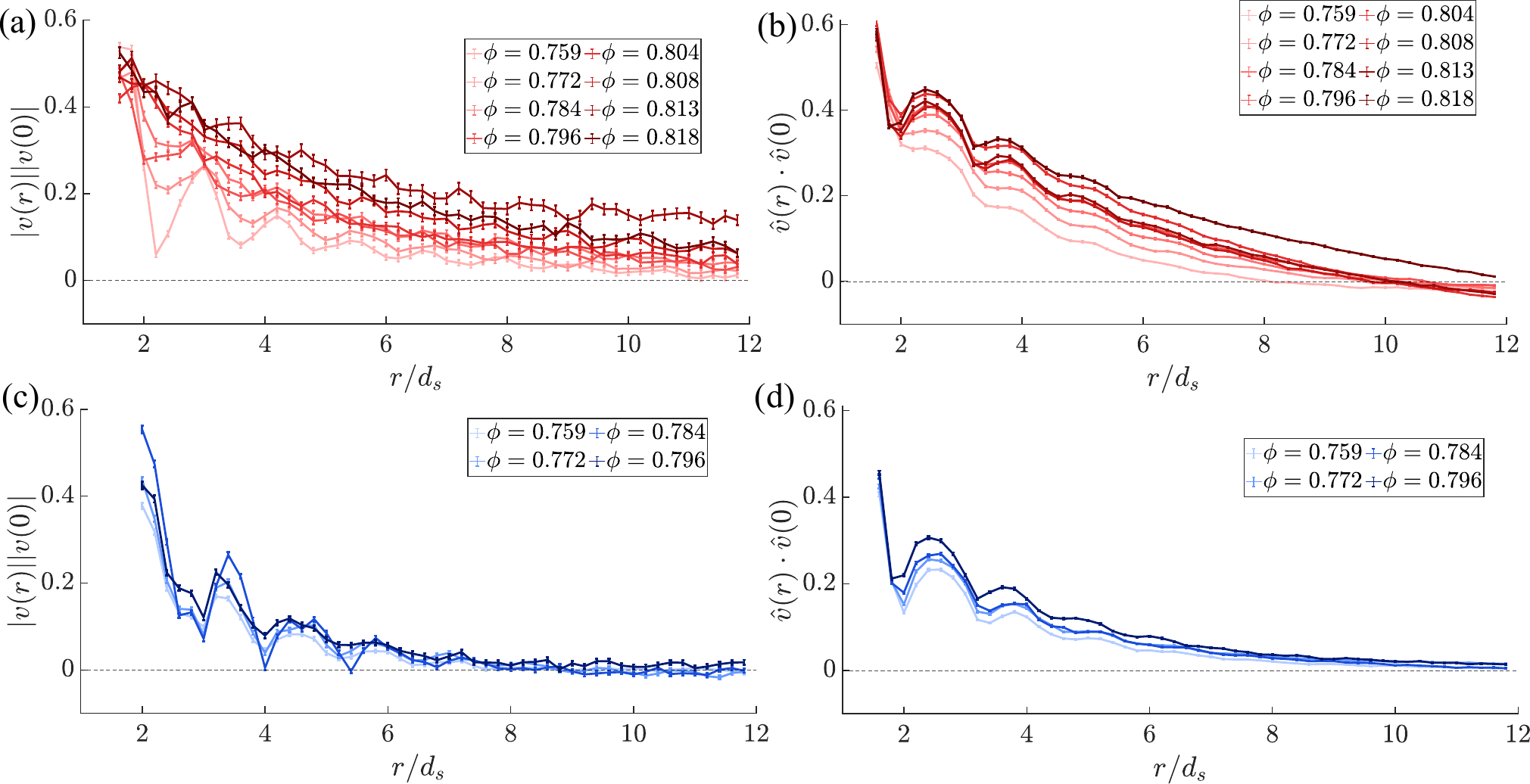}
    \caption{Spatial correlations of the velocity increase with the area fraction, $\phi$. Active mixture: (a)Velocity magnitude correlation $\Gamma_{|\mathbf{v}|}(r)$, and (b) Velocity direction correlation $\Gamma_{|\hat{v}|}(r, \Delta t)$.  Passive mixture: (c)Velocity magnitude correlation, and (d) Velocity direction correlation. }
    \label{fig:corr_all_lasp_pass}
\end{figure}

In Fig.\ref{fig:corr_all_lasp_pass}, we display the velocity magnitude and direction correlations for several area fractions for both active and passive mixtures. The correlation increases with the area fraction, $\phi$, and are longer in range than in the passive mixtures. 

\begin{figure}[h!]
    \centering
    \includegraphics[width=1\linewidth]{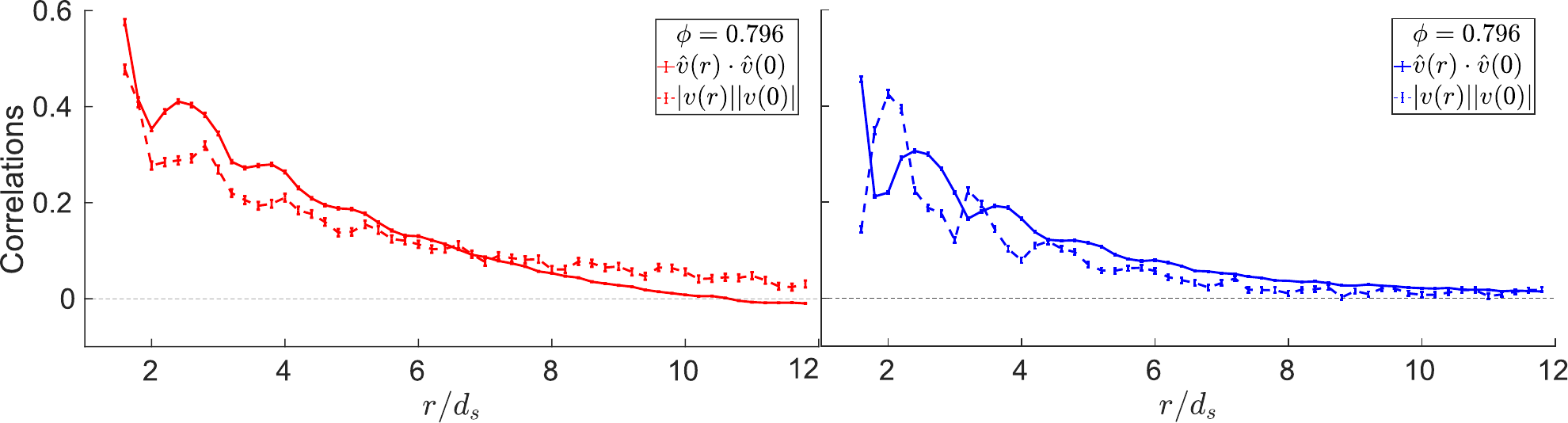}
    \caption{Spatial correlations (all-all) in the active system (left panel) and passive system (right panel) at $\phi = 0.796$ calculated at lag time $\Delta t = \tau_{\theta}$. Correlation in velocity magnitude (dashed) is higher than in the velocity direction (solid) for the active mixture, where as they are very similar in the passive mixture.}
    \label{fig:corr_325_both}
\end{figure}

For the active mixture, as we approach the glassy regime, the correlation in the magnitude of the velocity is higher than the correlation in the direction of the velocity (Fig.\ref{fig:corr_325_both}). 
\\

\begin{figure}[h!]
    \centering
    \includegraphics[width=0.8\linewidth]{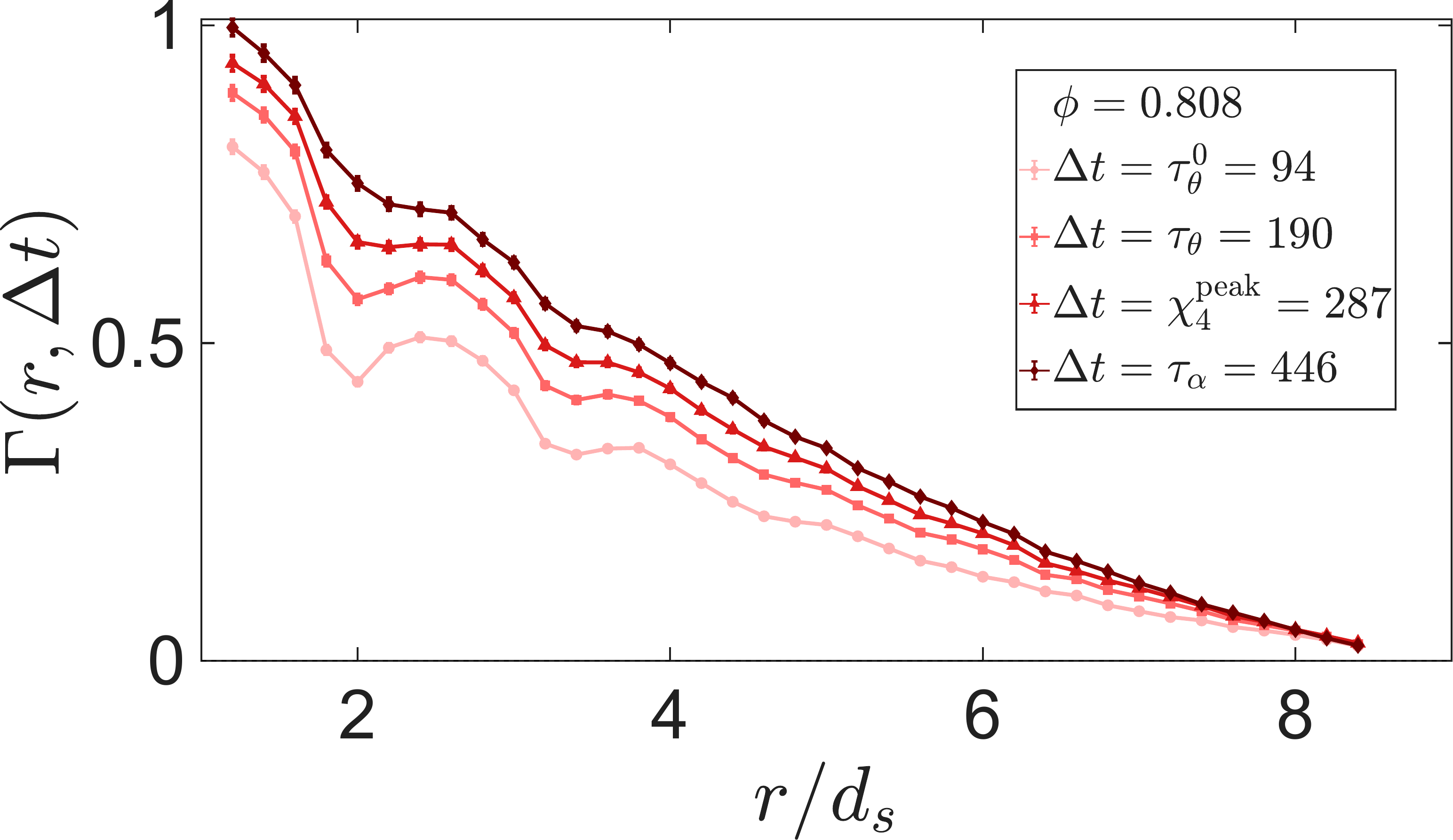}
    \caption{Spatial correlations (all-all) in Active system $\phi = 0.808$. Velocity correlation increases with the lag time. Here lag times chosen are relevant timescale (unit - no. of shakes) in the system.}
    \label{fig:velr_330}
\end{figure}

\newpage
In Fig. \ref{fig:velr_330}, we show the velocity correlations $\Gamma(r,\Delta t)$ at a high area fraction ($\phi = 0.808$) for several choices of the lag time, $\Delta t$.   The amplitude of correlations increase with the lag time. 

\subsection{Dynamical heterogeneity of large vs. small particles}
In the main text (Fig.3) we showed that in the active mixture, the small and large particles shared the same relaxation time, whereas in the passive system, the small particles have faster dynamics. Here we go beyond the mean relaxations to compare the dynamics heterogeneity in the two populations of particles in a mixture. 

\begin{figure}[h!]
    \centering
    \includegraphics[width=0.8\linewidth]{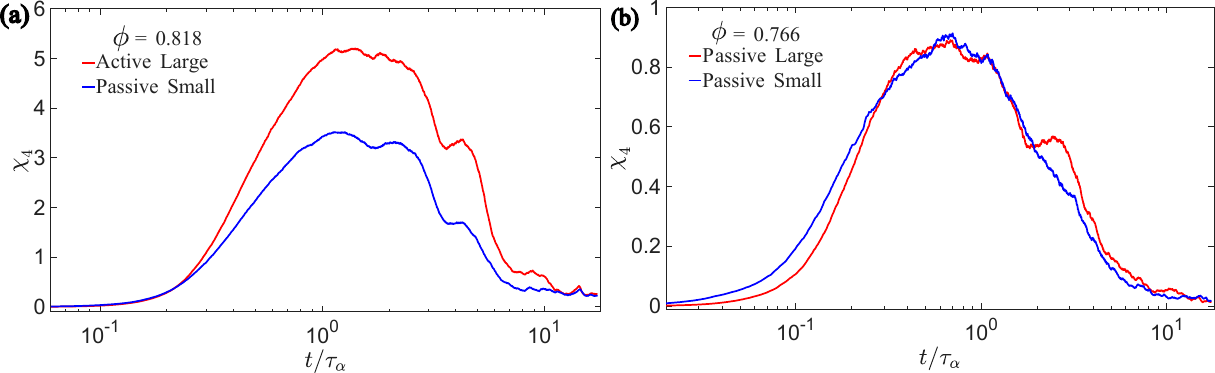}
    \caption{Heterogeneity among large and small particles. (a) In the active mixture $\chi_4$ in large particles and small particles tracks each other very well, down to the fluctuations, though they have different amplitude. (b) In the passive mixture the amplitude and peak frequency of the two species is the same, however the small fluctuations are different.}
    \label{fig:x4_comp}
\end{figure}

In Fig. \ref{fig:x4_comp} we plot $\chi_4$ against $t/\tau_\alpha$ for a pair of mixtures which have the same relaxation time, but now separate out the motions of the small and large particles. In both mixtures, $\chi_4$ peaks at the same timescale for both species.  
 In the active mixture (Fig. \ref{fig:x4_comp}(a)) $\chi_4$ for the large and small particles follow each other closely, with even the noise in the data matching each other. This is not the case in the passive mixture (Fig. \ref{fig:x4_comp}(b)), where, however, the amplitude of the peak is the same for the two species. 

\newpage

\end{document}